\documentclass[preprint,aps,floatfix,nofootinbib,preprintnumbers,amsmath, amssymb,showkeys]{revtex4-1}

\usepackage{epsf,epsfig,subfigure,graphicx,amsmath,amssymb}
\usepackage{color}
\usepackage{slashed}
\usepackage{amsmath}
\usepackage{mathtools}
\usepackage{accents}
\usepackage{color}
\usepackage{makecell}
\allowdisplaybreaks

\def\lsim{\mathrel{\rlap{\lower4pt\hbox{\hskip1pt$\sim$}}
		\raise1pt\hbox{$<$}}}         %less than or approx. symbol
\def\gsim{\mathrel{\rlap{\lower4pt\hbox{\hskip1pt$\sim$}}
		\raise1pt\hbox{$>$}}}         %greater than or approx. symbol

\begin{document}
	
	\title{\bf 	Quaternion Electromagnetism and \\
the Relation with 2-Spinor Formalism }

%	\author{
%		I.~K.~ Hong \footnote{hijko3@naver.com} and C.~S.~ Kim \footnote{cskim@yonsei.ac.kr}	}
%		
%	\affiliation{
%		Department of Physics and IPAP, Yonsei University, Seoul 120-749, Korea\\	}

%\author{I.~K.~ Hong$^1$\footnote{Email at: hijko3@naver.com}\, and C.~S.~ Kim$^{1,2}$\footnote{Email at: 
\author{I.~K.~ Hong$^1$\footnote{Email at: hijko3@yonsei.ac.kr}\, and C.~S.~ Kim$^{1,2}$\footnote{Email at: cskim@yonsei.ac.kr}\\[5mm]$^1$Department of Physics and IPAP, Yonsei University,  %\\[2mm]
Seoul 03722, Korea\\[5mm]$^2$Institute of High Energy Physics, Dongshin University,  %\\[2mm]
Naju 58245, Korea}%

	\vspace{1.0cm}
	\begin{abstract}
By using complex quaternion, which is the~system of quaternion representation extended to complex numbers, we show that the~laws of %physics of
electromagnetism can be expressed much more simply and concisely. We~also derive the~quaternion representation of rotations and boosts from the~spinor representation of Lorentz group.
It is suggested that the~imaginary ``$i$'' should be attached to the~spatial coordinates, and observe that the~complex conjugate of quaternion representation is exactly equal to parity inversion of all physical quantities in the~quaternion.
We~also show that using quaternion is directly linked to the two-spinor  formalism.
Finally, we discuss meanings of quaternion, octonion and sedenion in physics as n-fold rotation
	\end{abstract}


	\maketitle

	%%%%%%%%%%%%%%%%%%%%%%%%%%%%%%%%%%%%%%%%%%%%%%%%%%%%%%%%%%%%%%%%%%%
\section{Introduction}
There are several papers claiming that the~quaternion or the~octonion can be used to describe the~laws of classical electromagnetism in a simpler way \cite{1,2,22,3,4,5}.
However, they are mainly limited to describing Maxwell equations. Furthermore, the~meaning of quaternion and the~reasons   electromagnetic laws can be concisely described by them have not been well discussed up to now.
Here,~we list more diverse quaternion representations of the~relations in electromagnetism than previously known and we introduce a new simpler notation to express quaternions. The~proposed notation makes the~quaternion representation of electromagnetic relations look similar to the~differential-form representation of them. Moreover, the~classical electromagnetic mass density and the~complex Lagrangian can be newly defined and used to represent electromagnetic relations as quaternions.

It has been already well known that the~quaternion can describe the~Lorentz transformations of four vectors \cite{de1996quaternions}. We~here rederive the~quaternion representation of the~Lorentz boost and the~rotation, by using isomorphism between the~basis of quaternion and the~set of sigma matrices. Hence,~we~find that not only four vector quantities but also electromagnetic fields can be transformed simply in the~quaternion representation.
Starting from the~$4\times 4$ matrix representation of quaternion, we define a new complex electromagnetic field tensor. By~using~it, a complex energy--momentum stress tensor of electromagnetic fields and a complex Lagrangian can be nicely expressed. Interestingly, the~eigenvalues of the~complex energy--momentum stress tensor are the~classical electromagnetic mass density up to sign. To define complex tensors, we introduce a new spacetime index called ``tilde-spacetime index''.
Imaginary number $i$ is usually linked to time so that it can be regarded as imaginary time, but we   insist that it is more natural for $i$ to be linked to space. In our representation, we also find that the~complex conjugate of a quaternion is equal to the~quaternion consisting of the~physical quantities with parity~inversion.

The two-spinor formalism is known to be a spinor approach, which is useful to deal with the~general relativity \cite{penrose1960spinor, bain2000coordinate}. In the~formalism, all world-tensors can be changed to even-indexed spinors and there we   derive spinor descriptions of electromagnetism \cite{penrose1984spinors}.
We~here prove that the~quaternion representations including Maxwell's equations
are equivalent to the~spinor representations of electromagnetism.
We~also explain how spinors in two-spinor formalism are generally linked to the~quaternion.
Finally, we explore the~meaning of quaternion and more extended algebras such as octonion as n-fold rotation.

\section{Complex Quaternion}
Let us denote quaternions by characters with a lower dot such as $\d{q}$.
Quaternions are generally represented in the~form
\begin{eqnarray}
\d{q}=s+v_1 \mathbf{i}+v_2\mathbf{j}+v_3\mathbf{k} \label{1}
\end{eqnarray}
where $s, v_1, v_2, v_3$ are real numbers and $\mathbf{i}, \mathbf{j}, \mathbf{k}$ are the~units of quaternions which satisfy
\begin{eqnarray}
\mathbf{i}^2=\mathbf{j}^2=\mathbf{k}^2=-1,\quad \mathbf{i}\mathbf{j}=-\mathbf{j}\mathbf{i}=\mathbf{k},\quad \mathbf{j}\mathbf{k}=-\mathbf{k}\mathbf{j}=\mathbf{i},\quad \mathbf{k}\mathbf{i}=-\mathbf{i}\mathbf{k}=\mathbf{j}. \label{eq2}
\end{eqnarray}

Equation~(\ref{1}) consists of two parts, namely a ``scalar'' part $s$ and a ``quaternion vector'' part $v_1 \mathbf{i}+v_2\mathbf{j}+v_3\mathbf{k}$. If we denote the~quaternion vector part by $\vec{v}$, Equation  (\ref{1}) is written as
\begin{eqnarray}
\d{q}=s+\vec{v}.
\end{eqnarray}

All quaternion vectors, denoted by an over-arrow symbol  $\;\vec{}\;$,
can be interpreted as  coordinate vectors in $\mathbb{R}^3$. We~do      not distinguish between vectors and quaternion vectors in this paper.

If $\d{q}_1=a+\vec{A}$ and $\d{q}_2=b+\vec{B}$ are two quaternions, the~multiplication of the~quaternions can be described as	
\begin{eqnarray}
\d{q}_1 \d{q}_2 =(a+\vec{A})(b+\vec{B})=ab-\vec{A}\cdot\vec{B} +a\vec{B}+b\vec{A}+\vec{A} \times \vec{B}	, \label{2}
\end{eqnarray}
by applying Equation (\ref{eq2}), where $\vec{A}\cdot\vec{B}$ is the~dot product and $\vec{A} \times \vec{B}$ is the~cross product.
The~dot product and the~cross product, which are operations for three-dimensional vectors are used in quaternion vectors.

The~components of quaternions can be extended to complex numbers. We~call such a quaternion  ``complex quaternion''. The~general form of complex quaternion is
\begin{eqnarray}
\d{Q}=a+ib+\vec{c}+i\vec{d}~.
\end{eqnarray}
where $a,b$ and components of $\vec{c},\vec{d}$ are real numbers, and $i$ is a complex number $\sqrt{-1}$, which differs from the~quaternion unit $\mathbf{i}$.

We~denote the~operation of complex conjugation by a bar $\;\bar{}\;\;$, and the~complex conjugate of $\d{Q}$ is
\begin{eqnarray}
\bar{\d{Q}}=a-ib+\vec{c}-i\vec{d} ~.
\end{eqnarray}

For a quaternion vector $\vec{q}=q_1 \mathbf{i}+q_2 \mathbf{j}+q_3 \mathbf{k}$, the~exponential of $\vec{q}$ is defined by
\begin{eqnarray}
\exp (\vec{q})= e^{\vec{q}}\equiv 1+\d{q}+\frac{1}{2!} \d{q}^2+\frac{1}{3!}\d{q}^3... =\cos{|\vec{q}|}+i \, \frac{\vec{q}}{q}\sin{|\vec{q}|}~,
\end{eqnarray}
since $\d{q}^2=-|\vec{q}|^2$~ \cite{liu2003parameterization}.

%%%%%%%%%%%%%%%%%%%%%%%%%%%%%%%%%%%%%%%%%%
\section{Laws of Electromagnetism in the~Complex Quaternion Representation}
\vspace{-6pt}
\subsection{Electromagnetic Quantities} %Please confirm formatting of subsection titles.

We~use the~unit system which satisfies $\epsilon_0 =\mu_0 =c=1$ where $\epsilon_0$ is vacuum permittivity, $\mu_0$~is vacuum permeability and $c$ is speed of light.  The~sign conventions for the~Minkowski metric is $g_{\mu \nu}=\rm{diag}(1,-1,-1,-1)$.

In the~classical electromagnetism, the~density of electromagnetic field momentum $\vec{\mathfrak{p}}$ and the~density of electromagnetic field energy $\mathfrak{u}$ are defined by
\begin{eqnarray}
\vec{\mathfrak{p}} \equiv \vec{E} \times \vec{B},\qquad \qquad \mathfrak{u} \equiv \frac{1}{2} (|\vec{E}|^2+|\vec{B}|^2)~,  \label{8}
\end{eqnarray}
where $\vec{E}$ is an electric field and $\vec{B}$ is a magnetic field \cite{griffiths2005introduction, boyer1982classical}. In our unit system, the~electromagnetic momentum $\vec{\mathfrak{p}}\equiv \epsilon_0 \vec{E} \times \vec{B}$ (in SI units) is the~same as the~Poynting vector $\vec{S}\equiv \frac{1}{\mu_0} \vec{E}\times\vec{B}$ (in SI units).

We~define a complex Lagrangian $\mathfrak{L}$ and an  electromagnetic mass density $\mathfrak{m}$ by
\begin{eqnarray}
\mathfrak{L} \equiv \frac{1}{2} (|\vec{E}|^2-|\vec{B}|^2)+i \,\vec{E} \cdot \vec{B},  
\qquad
\mathfrak{m} \equiv \sqrt{ \mathfrak{u}^2 -\mathfrak{p}^2}= \sqrt{\frac{1}{4} (|\vec{E}|^2-|\vec{B}|^2)^2+(\vec{E} \cdot \vec{B})^2}~.   \label{LM}
\end{eqnarray}

The~electromagnetic mass density $\mathfrak{m}$ is defined from the~energy--momentum relation $m^2=u^2-|\vec{p}|^2$ where $(u, \vec{p})$ is four-momentum of a particle of mass $m$. The~meaning of $\mathfrak{m}$ should be investigated
more in detail; however, it is not discussed here.
Comparing $\mathfrak{L}$ and $\mathfrak{m}$, we can see that
\begin{eqnarray}
\mathfrak{m} =\sqrt{\mathfrak{L} \bar{\mathfrak{L}}}~.
\end{eqnarray}

\subsection{Complex Quaternion Representations of Electromagnetic Relations}

Let us define a few physical quantities in the~form of complex quaternion,
\begin{eqnarray}
\begin{matrix*}[l]
\d{u} \equiv \gamma+i\,\gamma \vec{v},   &\qquad \qquad \;
\d{A} \equiv V+ i\vec{A},   \\
\d{F} \equiv i\vec{E} -\vec{B}, &\qquad \qquad \;
\d{J} \equiv \rho+i\vec{J},  \\
\d{p} \equiv  \mathfrak{u} + i\vec{\mathfrak{p}}, &\qquad \qquad \;
\d{f} \equiv \vec{J} \cdot \vec{E}+ i ( \rho \vec{E} + \vec{J} \times \vec{B} )~,
\end{matrix*}
\end{eqnarray}
where $\gamma$ is $1/\sqrt{1-v^2}$ for the~velocity $v$, $V$ is the~electric potential, $\vec{A}$ is the~vector potential, $\vec{E}$ is the~electric field, $\vec{B}$ is the~magnetic field, $\rho$ is the~charge density, and $\vec{J}$ is the~electric current density.
$\d{F}$~is just a quaternion vector and the~terms in $\d{p}$ are defined in Equation       (\ref{8}). $\d{J}$ is equal to $\rho_0 \d{u}$ where $\rho_0$ is the~proper charge density, which is the~density in the~rest system of the~charge. The~scalar part of $\d{f}$ is the~rate of work done by electric field on the~charge and the~vector part is the~Lorentz force. 

We~define a quaternion differential operator by
\begin{eqnarray}
\d{d} \equiv  \frac{\partial}{\partial t}-i\nabla~,
\end{eqnarray}
where $t$ is the~time and $\nabla=\partial_x \mathbf{i} + \partial_y\mathbf{j} +\partial_z \mathbf{k}$ is the~vector differential operator in the~three-dimensional Cartesian coordinate system.
\\

The~relations in electromagnetism can be described in the~complex quaternion form simply as~follows:
\begin{eqnarray}
\begin{matrix*}[l]

1)\; \d{d} \bar{\d{d}}= \Box^2 \qquad &\text{(d'Alembert Operator)} \\ 
2)\; \d{A'}=\d{A}+\d{d} \lambda \qquad &\text{(Gauge Transformation)} \\
3)\; \d{d} \bar{\d{A}} =\d{F} \qquad &\text{(Field Strength from Gauge Field)}\\
4)\; \d{d} \bar{\d{F}}=	\d{J} \, ( = \d{d} \bar{\d{d}} \d{A}) \qquad &\text{(Electromagnetic Current, Maxwell Equations)} \\
5)\; \d{d} \bar{\d{J}}=\d{d} \bar{\d{d}} \d{F}=	\Box^2 \d{F} \qquad &\text{(Electromagnetic Wave Equation with Source)}\\
6)\;  \d{F} \d{J} = \d{f} +\d{l} \qquad &\text{(Lorentz Force)} \\
7)\; \d{F}(\d{d}\bar{\d{F}})=(\d{F}\d{d})\bar{\d{F}} (= \d{F} \d{J}) \qquad &\text{(Formula with Quaternion Differential Operator)} \\
8)\;\frac{1}{2} \d{F} \bar{\d{F}}= \d{p} \qquad &\text{(Electromagnetic Energy--Momentum)}\\
9) \; \d{d} \bar{\d{p}} =  \frac{1}{2} \left[ (\d{d} \bar{\d{F}} ) \d{F}+\bar{\d{F}} (\bar{\d{d}} \d{F}) \right ] + i (\bar{\d{F}}\cdot \nabla ) \d{F} &\text{(Conservation of Electromagnetic Energy--Momentum)} \\
10)\; \frac{1}{2} \d{F} \d{F}= \mathfrak{L} \qquad &\text{(Euclidean Lagrangian of Electromagnetic Fields)}\\
11)\; \d{p}\bar{\d{p}}= \mathfrak{L} \bar{\mathfrak{L}} =  \mathfrak{m}^2 \qquad &\text{(Electromagnetic Mass Density)}\\
\end{matrix*}  \label{9}
\end{eqnarray}
where $\d{l}=i\vec{J} \cdot \vec{B} + (-\rho \vec{B}- \vec{E} \times \vec{J} )$.

We~can check all quaternion relations by expanding multiplications of quaternions using Equation~(\ref{2}). Some expansions are proven in Appendix \ref{appA}.
Relations (1), (3) and (4)    are   already well known in quaternion forms, but the~others are not   well mentioned  thus  far.
Each~quaternion equation in Equation       (\ref{9})  contains several relations, which are known in classical electromagnetism.
\\

Let us discuss in more detail each relation in Equation (\ref{9}).

	1) $\d{d} \bar{\d{d}}= \Box^2$ is the~d'Alembert operator.

	2) $\d{A}'=\d{A}+\d{d} \lambda$ describes the~gauge transformation of gauge fields.
	\begin{eqnarray}
	V'=V+\frac{\partial \lambda}{\partial t}, \qquad \quad \vec{A}'=\vec{A}+\nabla \lambda.
	\end{eqnarray}
	
	3) $\d{d} \bar{\d{A}} = \d{F} $ contains three relations. One is Lorentz gauge condition and the~others are the~relations between  fields strength and gauge fields, as shown in Equation~(\ref{dA}),
	
	\begin{eqnarray}
	&& 	\qquad \quad  \frac{\partial V}{\partial t} +\nabla \cdot \vec{A} =0 \\
	&&\vec{E}=-\nabla V+\frac{\partial \vec{A}}{\partial t}, \qquad \quad \vec{B}=\nabla \times \vec{A}.
	\end{eqnarray}
	
	4) $ \d{d} \bar{\d{F}}=	\d{J} \, ( = \d{d} \bar{\d{d}} \d{A})$ contains all  four  Maxwell's equations in Equation (\ref{dF}),
	\begin{eqnarray}
	&&\nabla \cdot \vec{E} = \rho, \qquad
	\nabla \times \vec{B} = J + \frac{\partial \vec{E}}{ \partial t},  \label{10} \\
	&&\nabla \cdot \vec{B}= 0, \qquad
	\;\;\nabla \times \vec{E}= -\frac{\partial \vec{B}} {\partial t}. \label{11}
	\end{eqnarray}
	
	It can be the~wave equations of gauge fields with sources in the~Lorentz gauge,
	\begin{eqnarray}
	\Box^2 \vec{V}= \rho, \qquad \quad \Box^2 \vec{A}= \vec{J}.
	\end{eqnarray}
	
	5)  $ \d{d} \bar{\d{J}}=\d{d} \bar{\d{d}} \d{F}=	\Box^2 \d{F} $ contains the~charge conservation relation and the~wave equations of $\vec{E}$ and $\vec{B}$ fields   in Equation (\ref{dJ}),
	\begin{eqnarray}
	&& \qquad \qquad \quad  \frac{\partial \rho}{\partial t}+ \nabla \cdot \vec{J} =0, \\
	&&\Box^2 \vec{E}= -\nabla \rho+ \frac{\partial J}{\partial t}, \qquad \quad \Box^2 \vec{B}=\nabla \times \vec{J}.
	\end{eqnarray}
	
	Those can be  derived from taking $\d{d}$ operation \vspace{2pt} on the~both side of Relation     (4) in Equation~(\ref{9}).
	
	6) $\d{F} \d{J}$ includes the~Lorentz \vspace{2pt} force term $\rho \vec{E} + \vec{J} \times \vec{B}$ and the~work done by electromagnetic fields term $\vec{J}\cdot \vec{E}$. However, the~meaning of \vspace{2pt} $\d{l}=i  \vec{J} \cdot \vec{B} + (-\rho \vec{B}- \vec{E} \times \vec{J} )$ is not yet well known.
	
	7) We~have found that $(\d{F}\d{d})\bar{\d{F}}$ \vspace{2pt} is equal to $\d{F}(\d{d}\bar{\d{F}})(= \d{F} \d{J})$ where $(\d{F}\d{d})$ is the quaternion differential operator. The~proof of this is given in Appendix \ref{fd}.
	
	8) $\;\frac{1}{2} \d{F} \bar{\d{F}}= \d{p}$ is the~quaternion representation of electromagnetic energy and momentum. It can be easily verified, by expanding the~left side, that
	\begin{eqnarray}
	\mathfrak{u}=\frac{1}{2} (|\vec{E}|^2+|\vec{B}|^2),\qquad \quad \vec{\mathfrak{p}}=\vec{E} \times \vec{B}. \label{17}
	\end{eqnarray}
	
	9) It can be guessed that $\d{d} \bar{\d{p}} \sim \d{f} $ from the~analogy with the~force-momentum relation
	$ \frac{D p_\lambda}{D \tau}=f_\lambda =q u^\mu F_{\mu \lambda}$, where $f_\lambda$ is the~four-force, $D$ is the~covariant derivative, $\tau$ is the~proper time, $q$ is the~electric charge, $U^\mu$ is the~four-velocity, and $F_{\mu \lambda}$ is the~electromagnetic tensor, which is the relation of the~four-force acting to a charged particle situated in electromagnetic fields.
	$\d{d}\bar{\d{p}}$ is expanded as
	\begin{eqnarray}
	\d{d} \bar{\d{p}}=(\partial_{t}-i\nabla )(\mathfrak{u}-i\mathfrak{p})=(\partial_{t} \mathfrak{u} +\nabla \cdot \mathfrak{p}) - i \,(\partial_{t} \mathfrak{p}+\nabla \mathfrak{u})  - (\nabla \times \mathfrak{p}). \label{dp}
	\end{eqnarray}
	Substituting  Equation (\ref{17}) into Equation        (\ref{dp}), we get
	\begin{eqnarray}
	\partial_{t} \mathfrak{u} +\nabla \cdot \mathfrak{p}&&\;=-\vec{J} \cdot \vec{E} \label{24} \\
	-(\partial_{t} \mathfrak{p}+\nabla \mathfrak{u})&&\;= (\rho \vec{E} +\vec{J} \times \vec{B})-(\nabla \cdot \vec{E}) E_i -(\vec{E}\cdot \nabla) E_i-(\nabla \cdot \vec{B}) \vec{B}_i -(\vec{B}\cdot \nabla) B_i \label{25}\\
	\nabla \times \mathfrak{p}&&\;= (-\rho \vec{B} + \vec{J} \times \vec{E}) - \vec{E} \times \partial_{t} \vec{E} - \vec{B} \times \partial_{t} \vec{B} +\vec{E} \nabla \vec{B} - \vec{B} \nabla \vec{E},  \label{26}
	\end{eqnarray}
	where $(\vec{A}\nabla \vec{B})_i\equiv A_j(\nabla_i B_j)$ for  vector fields $\vec{A}$ and $\vec{B}$.
	
	Equation~(\ref{24}) is the~work--energy relation in electromagnetism.
	Equation~(\ref{25}) can be rearranged~as
	\begin{eqnarray}
	\mathfrak{f}= (\overleftrightarrow{\nabla} \cdot \mathbb{T}) -\frac{ \partial \mathfrak{p}}{\partial t },
	\end{eqnarray}
	where
	\begin{eqnarray}
	(\mathbb{T})_{ij}=(E_i E_j-\frac{1}{2} \delta_{ij}|\vec{E}|^2)+(B_i B_j -\frac{1}{2} \delta_{ij} |\vec{B}|^2) \label{st}
	\end{eqnarray} 
	is the~Maxwell stress tensor
	and
	\begin{eqnarray}
	(\overleftrightarrow{\nabla} \cdot \mathbb{T})_i \equiv (\nabla \cdot \vec{E}) E_i +(\vec{E}\cdot \nabla) E_i+(\nabla \cdot \vec{B}) \vec{B}_i +(\vec{B}\cdot \nabla) B_i-\frac{1}{2}\nabla_i (|\vec{E}|^2+|\vec{B}|^2). \label{tt}
	\end{eqnarray}
	
	Equation~(\ref{26}) is not a well-known relation. The~proof of the~expansion is given in Appendix \ref{ntp}.
	By looking at Equations (\ref{24})--(\ref{26}), {we can observe that} it is difficult to find a simple quaternion formula such as $\d{d} \bar{\d{p}} = \d{f}+\d{l} $.
	The~exact formula of $\d{d}\bar{\d{p}}$ is obtained as
	\begin{eqnarray}
	\d{d}\bar{\d{p}}&& = \frac{1}{2} \left[ (\d{d} \bar{\d{F}} ) \d{F}+\bar{\d{F}} (\bar{\d{d}} \d{F}) \right ] + i \, (\bar{\d{F}}\cdot \nabla ) \d{F}. \label{22}
	\end{eqnarray}
	
	%through another analysis.
	{The~proof is given in Appendix \vspace{2pt} \ref{qq}}.
	
	10) $\frac{1}{2} \d{F} \d{F}= \mathfrak{L}$ is \vspace{2pt} the~relation between the~complex Lagrangian and electromagnetic fields.
	The~complex Lagrangian $\mathfrak{L}$ is defined as $ \frac{1}{2} (|\vec{E}|^2-|\vec{B}|^2)+i\,\vec{E} \cdot \vec{B}  $. \vspace{2pt} This is, in fact, the~Euclidean Lagrangian including topological term \cite{polyakov1987gauge, preskill1984magnetic}.
	The~real part $ \frac{1}{2} (|\vec{E}|^2-|\vec{B}|^2) $ is the~Lagrangian \vspace{2pt} of electromagnetic fields $\frac{1}{4}F_{\mu\nu}F^{\mu\nu}$, where $F_{\mu \nu}=A_{ [\mu} \partial_{\nu]}$ for $U(1)$ gauge field $A_\mu$. The~variation \vspace{2pt} of this part gives the~first two Maxwell's equations in Equation (\ref{10}). The~complex part $\vec{E} \cdot \vec{B}$ is $\frac{1}{4}F_{\mu\nu}\; {}^*F^{\mu\nu}$, which is the~topological term of gauge fields where ${}^*F^{\mu\nu}$ is Hodge dual of $F^{\mu\nu}$. Its variation gives the~other two Maxwell's equations in Equation (\ref{11}).
	
	11) $\d{p}\bar{\d{p}}= \mathfrak{L} \bar{\mathfrak{L}} =  \mathfrak{m}^2$ is a Lorentz invariant and a gauge invariant quantity.

\section{Lorentz Transformation in the~Complex Quaternion Representation  \label{LTQ}}

For a quaternion basis \vspace{2pt} $\{ 1, \mathbf{i},  \mathbf{j}, \mathbf{k} \}$,
the algebra of $\{ 1,i \, \mathbf{i}, i \, \mathbf{j},i \, \mathbf{k} \} $ is isomorphic to the~algebra of sigma matrices  $\{ \sigma^0, \sigma^1, \sigma^2, \sigma^3 \}$, where $\sigma^0$ is $2 \times 2$ identity matrix and $\sigma^1, \sigma^2, \sigma^3$ are Pauli matrices,
\begin{equation}
\sigma^0=\begin{pmatrix} 1&0 \\0&1 \end{pmatrix}, \quad
\sigma^1=\begin{pmatrix} 0&1 \\1&0 \end{pmatrix},   \quad
\sigma^2=\begin{pmatrix} 0&-i \\i&0 \end{pmatrix},   \quad
\sigma^3=\begin{pmatrix} 1&0 \\0&-1 \end{pmatrix}. 
\end{equation}

It means that complex quaternions that have the~form $\d{q}=q_{0}+i \vec{q}$ are isomorphic to $q_{\mu} \sigma^\mu$ where $q_{\mu}=(q_{0}, \vec{q})=(q_{0}, q_1,q_2,q_3)$ \vspace{2pt} and $\sigma^\mu=(\sigma^0, \sigma^1, \sigma^2, \sigma^3 )$.

We~can get the~quaternion representation of Lorentz transformation by using isomorphism given above and the~spinor representation of the~Lorentz group.
Let us denote by $S[\Lambda]$ the~spinor representation of the~Lorentz group,  which acts on Dirac spinor $\psi(x)$. Then, Dirac spinor transforms as $\psi(x) \rightarrow S[\Lambda] \psi(\Lambda^{-1} x)$ under a Lorentz transformation $x \rightarrow x'=\Lambda x$.

In the~chiral representation of the~Clifford algebra, the~spinor representation of rotations $S[\Lambda_{rot}]$ and boosts $S[\Lambda_{boost}]$ are
\begin{eqnarray}
S[\Lambda_{rot}] =
\begin{pmatrix}
e^{+i\vec{\phi} \cdot \vec{\sigma}/2} & 0 \\
0 & e^{+i\vec{\phi} \cdot \vec{\sigma}/2}
\end{pmatrix}
, \qquad
S[\Lambda_{boost}] =
\begin{pmatrix}
e^{+\vec{\eta} \cdot \vec{\sigma}/2} & 0 \\
0 & e^{-\vec{\eta} \cdot \vec{\sigma}/2}
\end{pmatrix},
\end{eqnarray}
where $\vec{\phi}= \phi \hat{\phi} $, $\vec{\eta}=\hat{v}\tanh^{-1} |\vec{v}| $,  $\phi$ is the~rotation angle, $\hat{\phi}$ is the~unit vector of rotation axis, $\vec{v}$ is the~boost velocity, and $\hat{v}$ is the~unit vector of boost velocity.

Since it is known \cite{tong2007quantum} that
\begin{eqnarray}
S[\Lambda]^{-1}\gamma^\mu S[\Lambda]= \Lambda^\mu_{\;\; \nu} \gamma^\nu,  
\end{eqnarray}
the following relation also holds:
\begin{equation}
S[\Lambda]^{-1} V_\mu \gamma^\mu S[\Lambda]= V_\mu \Lambda^\mu_{\;\; \nu} \gamma^\nu \label{30}
\end{equation}
for any four-vector $V^\mu$.

The~components of Equation        (\ref{30}) are
\begin{eqnarray}
\begin{pmatrix}
0 & e^{-i\vec{\phi} \cdot \vec{\sigma}/2}\; V_\mu\sigma^\mu \;e^{+i\vec{\phi} \cdot \vec{\sigma}/2}\\
e^{-i\vec{\phi} \cdot \vec{\sigma}/2} \; V_\mu \bar{\sigma}^\mu \; e^{+i\vec{\phi} \cdot \vec{\sigma}/2} & 0
\end{pmatrix}= \begin{pmatrix}
0 &  V_\mu \Lambda^\mu_{\;\; \nu} \; \sigma^\nu\\
V_\mu \Lambda^\mu_{\;\; \nu}\;  \bar{\sigma}^\nu & 0
\end{pmatrix},   \label{28}
\end{eqnarray}
\begin{eqnarray}
\begin{pmatrix}
0 & e^{-\vec{\eta} \cdot \vec{\sigma}/2} \; V_\mu\sigma^\mu \; e^{-\vec{\eta} \cdot \vec{\sigma}/2}\\
e^{+\vec{\eta} \cdot \vec{\sigma}/2} \; V_\mu \bar{\sigma}^\mu \; e^{+\vec{\eta} \cdot \vec{\sigma}/2} & 0
\end{pmatrix}= \begin{pmatrix}
0 &  V_\mu \Lambda^\mu_{\;\; \nu} \; \sigma^\nu\\
V_\mu \Lambda^\mu_{\;\; \nu} \; \bar{\sigma}^\nu & 0
\end{pmatrix},   \label{29}
\end{eqnarray}
where $\bar{\sigma}^\mu=(\sigma^0, -\sigma^1, -\sigma^2, -\sigma^3 )$. This represents the~quaternion Lorentz transformation for the~form $q_0+i\vec{q}$, \vspace{2pt} since $\d{q}=q_0+i \vec{q} \;\; \sim  \;\;  q_{\mu} \sigma^\mu$.

Let us define Lorentz transformation factor $\zeta (\phi, \eta)$ by
\begin{eqnarray}
\zeta (\phi, \eta) \equiv  e^{+\frac{1}{2}\vec{\phi}} e^{-\frac{1}{2}i\vec{\eta} } = (\cos{\frac{\phi}{2}}+\hat{\phi} \sin{\frac{\phi}{2}} ) (\cosh \frac{\eta}{2}-i \, \hat{\eta }\sinh\frac{\eta}{2} ), \label{31}
\end{eqnarray}
where  $\cosh \eta = \gamma$,  $\sinh \eta= \gamma v$.
Since \vspace{2pt} $(\cosh \frac{\eta}{2}+i \, \hat{\eta }\sinh\frac{\eta}{2})= \gamma+i\, \gamma \vec{v}  $ is the~quaternion velocity $\d{u}(\vec{v})$ of a boosted frame with a boost velocity $\vec{v}$, Equation~(\ref{31}) can be rewritten as
\begin{eqnarray}
\zeta(\phi, \eta)=\d{R}(\vec{\phi})\bar{\d{u}}(\vec{v}),
\end{eqnarray}
where  $\d{R}(\vec{\phi})\equiv (\cos{\frac{\phi}{2}}+\hat{\phi} \sin{\frac{\phi}{2}})$.
The~inverse of $\zeta (\phi, \eta)$ and its complex conjugate are defined as
\begin{equation}
\zeta^{-1} (\phi, \eta) = e^{+\frac{1}{2}i\vec{\eta} } e^{-\frac{1}{2}\vec{\phi}}, \qquad \overline{\zeta^{-1}} (\phi, \eta) = e^{-\frac{1}{2}i\vec{\eta} } e^{-\frac{1}{2}\vec{\phi}}.
\end{equation}

From Equations~(\ref{28}) and~(\ref{29}), the~Lorentz transformations of a quaternion  that has the~form \mbox{$\d{V}=V^0+i\vec{V}$} is written as
\begin{eqnarray}
\d{V}'= \zeta \d{V} \overline{\zeta^{-1}}.
\end{eqnarray}

Therefore, the~Lorentz transformations of a quaternion gauge field $\d{A}$ and a quaternion strength field $\d{F}$~are
\begin{eqnarray}
\d{A}' = && \zeta \d{A} \overline{\zeta^{-1}}, \\
\d{F}'= &&\d{d'} \bar{\d{A'}}=  \zeta \d{d} \overline{\zeta^{-1}}  \bar{\zeta} \d{A} \zeta^{-1} = \zeta \d{F} \zeta^{-1}.
\end{eqnarray}

As an example, if we boost a frame with a speed $v$ along $x$ axis,
then
\begin{eqnarray}
\zeta &&= \d{u}(\vec{v})=\cosh \frac{\eta}{2}+i\, \hat{\eta }\sinh\frac{\eta}{2}=\gamma+ i\, \gamma v \mathbf{i} \nonumber\\
\d{A}'&&= \zeta \d{A} \overline{\zeta^{-1}}=(\gamma- i\, \gamma \vec{v}) \d{A}(\gamma- i\,\gamma \vec{v} ) \nonumber
\\&&= \gamma (V -A_1 v)+ i \,(\gamma (A_1 - V v) \mathbf{i} +  A_2  \mathbf{i} +  A_3  \mathbf{k}),
\\
\d{F}'&&=\zeta \d{F} \zeta^{-1}=(\gamma- i\,\gamma \vec{v} ) \d{F}(\gamma+i\,\gamma \vec{v} )  \nonumber \\
&& =i \, (E_1  \mathbf{i}+ \!\gamma (E_2 - B_3 v)\mathbf{j}+ \! \gamma (E_3 + B_2 v)\mathbf{k}) \!-\!(B_1 \mathbf{i} +\!\gamma(B_2  + E_3 v)\mathbf{j}+\!\gamma (B_3  - E_2 v )\mathbf{k} ),
\end{eqnarray}
which is a very efficient representation in computing rotations and boosts.

\section{The~Role of Complex Number ``$i$'' in Complex Quaternions }
\vspace{-6pt}
\subsection{Complex Space and Real Time}

In this  section, we explain that it is more natural to attach imaginary number $i$ to the~spatial coordinates rather than to the~time coordinate.
The~infinitesimal version of the~Lorentz transformation in one dimension is
\begin{eqnarray}
dt'=\gamma (dt -v dx ),   \qquad   dx'=\gamma (dx -v dt ).
\end{eqnarray}

This can be manipulated to
\begin{eqnarray}
dt'&&=\gamma (dt -v dx )   \nonumber  \\&&= \frac{1}{\sqrt{1+(\frac{i\;dx_s' \i}{dt_s'})^2}}(dx-vdt)=\frac{1}{\sqrt{(dt_s'^2+(i\;dx_s')^2)}} (dt dt_s'   +(i\,dx ) (i\,dx_s' )),   \label{33}\\  \nonumber \\
i\,dx'&&=i \,\gamma (dx -v dt ) \nonumber
\\&&= \frac{1}{\sqrt{1+(\frac{i\,dx_s' }{dt_s'})^2}}(i\,dx -i \,v dt)= \frac{1}{\sqrt{(dt_s'^2+(i\,dx_s' )^2)}} ( (i\,dx ) dt_s'- dt (i\,dx_s' ) ), \label{34}
\end{eqnarray}
where $v=dx'_s /dt'_s$ is a boost velocity, $dx'_s$ is an infinitesimal displacement of the~moving frame and $dt'_s$ is an infinitesimal time it takes for the~frame to move along the~displacement.

If we put imaginary number ``$i$'' to the~spatial coordinate as  Equations (\ref{33}) and (\ref{34}), the~Lorentz transformation can be seen as a kind of rotation,
\begin{eqnarray}
\; dt =r \cos \alpha, &&\;i\,dx =r \sin \alpha, \nonumber \\
\;  dt_s =r \cos \beta,&& \;  i\,dx'_s =r \sin \beta, \nonumber \\
\rightarrow \;\;	dt'=r \cos (\alpha-\beta), &&\; i\,dx'  =r \sin (\alpha-\beta),
\end{eqnarray}
for  pure imaginary angles \vspace{2pt} $\alpha$, $\beta$ and $r=\sqrt{(dt_s'^2+(i\,dx_s' )^2)}  $.

In contrast, if we put $i$ to the~time coordinate rather than to the~spatial coordinate, then
\begin{eqnarray}
\; i\,dt =r \cos \alpha, &&\;dx =r \sin \alpha, \nonumber \\
\;  i\,dt_s =r \cos \beta, && \;  dx'_s =r \sin \beta \nonumber \\
\rightarrow \;\; i\,dt'\neq r \cos (\alpha-\beta), &&\; dx'  \neq r \sin (\alpha-\beta),
\end{eqnarray}
which means that  Equations (\ref{33}) and (\ref{34}) cannot be regarded as a kind of rotation.

\subsection{Parity Inversion and Conjugate of $i$ }

All physical quantities that are located in the~real part of \vspace{2pt} quaternions, such as $\rho,\; V,\; \vec{B},\;$ \mbox{$\frac{1}{2}(E^2-B^2) $,} etc., \vspace{2pt} do not change signs under parity inversion; and all physical quantities that are located in the~imaginary part of quaternions, such as $\vec{A}, \;\vec{E}, \;\vec{J}, \;\vec{E}\cdot \vec{B} $, etc.,
change signs under parity inversion. This~means that the~operation of complex conjugation on a quaternion corresponds to the~parity inversion of the~physical quantities in the~quaternion representation.     The~reason is related to tilde-spacetime indices, which  are defined in Sections \ref{ts} and \ref{RS}.

All quantities in the~imaginary part may be regarded as ``imaginary quantities', not just as ``real  quantities placed in the~imaginary part'', i.e.   imaginary space, imaginary momentum, imaginary electric field, etc. It is the~same as replacing length units, such as  ``$\rm{meter}$'', with imaginary length unit such as ``$i\;\rm{meter} $''.

\section{ Complex Electromagnetic Tensor Related to Quaternion and  Electromagnetic Laws}
\vspace{-6pt}
\subsection{Electromagnetic Tensor with Tilde-Spacetime Index \label{ts}}
For a vector $\vec{b}=(b_1,b_2,b_3)$, let us define ``vector matrix of $\vec{b}$'' as
\begin{eqnarray}
\mathbf{b} = \begin{pmatrix}
b_{1}\\
b_{2} \\
b_{3}
\end{pmatrix},
\end{eqnarray}
and the~vector matrix by $\#$ notation as
\begin{eqnarray}
\mathbf{b}^\#= -\epsilon_{ijk}b^k=
\begin{pmatrix}
0 & -b_3  & b_2 \\
b_3 & 0   & -b_1 \\
-b_2 & b_1   & 0
\end{pmatrix},
\end{eqnarray}
where $\epsilon_{ijk}$ are the~Levi--Civita symbols.

Then, the~electromagnetic tensor $F^{\mu\nu}$ can be represented as
\begin{eqnarray}
F^{\mu \nu}= \partial^\mu A^\nu-\partial^\nu A^\mu=\begin{pmatrix}
0 & -E_1 & -E_2 & -E_3 \\
E_1 & 0 & -B_3 & B_2 \\
E_2 & B_3 &  0  & -B_1 \\
E_3 & -B_2 & B_1 & 0  &
\end{pmatrix}
=\begin{pmatrix}
0 & -\mathbf{E}^t \\
\mathbf{E} & \mathbf{B}^\#
\end{pmatrix},  \label{F}
\end{eqnarray}
where $\mathbf{E}$, $\mathbf{B}$ are vector matrix of $\vec{E}$, $\vec{B}$ and superscript ${\mathbf{E}}^t$ means the~transpose of a matrix $\mathbf{E}$.
The~dual tensor can be represented as
\begin{eqnarray}
G^{\mu \nu}=\frac{1}{2} \epsilon^{\mu\nu\rho\sigma}F_{\rho \sigma}=\begin{pmatrix}
0 & -\mathbf{B}^t \\
\mathbf{B} & -\mathbf{E}^\#
\end{pmatrix},
\end{eqnarray}
{where $\epsilon^{\mu\nu\rho\sigma}$ is the~rank-4 Levi--Civita symbol with the~sign convention $\epsilon^{0123}=+1$.}

Now, we define tensor indices with tilde \vspace{2pt} such as ``${\tilde{\mu}\tilde{\nu}\tilde{\rho}..}$'', called ``tilde-spacetime indices''.
$O^{\tilde{\mu}}$ and $O_{\tilde{\mu}}$ for any $O^{\mu}=(O^0,O^1,O^2,O^3)$ and
$O_{\mu}=(O_0,O_1,O_2,O_3)$ are defined as
\begin{eqnarray}
O^{\tilde{\mu}}= (O^0,i \,O^1,  i \,O^2,  i \,O^3 ), \qquad O_{\tilde{\mu}}= (O_0,-i \,O_1,  -i \,O_2,-i \,O_3).
\end{eqnarray}
Then, the~components of $O^{\tilde{\mu}}$ and $O_{\tilde{\mu}}$ become equal, \vspace{2pt} since
$O_0=O^0,\;O_1=-O^1,\;O_2=-O^2,O_3=-O^3$ in Minkowski metric.
As an example, \vspace{2pt}  $\bar{\sigma}^{\tilde{\mu}}$ is $(\sigma^0,-i \,\sigma^1,  -i \,\sigma^2,  -i \,\sigma^3)$. Since this is isomorphic to the~quaternion basis $( 1, \mathbf{i},  \mathbf{j}, \mathbf{k} )$, \vspace{2pt} we can rewrite  $\bar{\sigma}^{\tilde{\mu}}$ as a quaternion basis $\hat{q}^{\tilde{\mu}}$ so that $A_{\mu} \bar{\sigma}^\mu=A_{\tilde{\mu}} \hat{q}^{\tilde{\mu}}$.

Generally speaking, the~way to convert a quantity with multiple spacetime indices to the~quantity with multiple tilde-spacetime indices is multiplying with or dividing by imaginary number $i$ when each spacetime index has the~value 1, 2 or 3.
As an example, the~Minkowski metric with tilde indices is
$g_{\tilde{\mu}\tilde{\nu}}=(1,1,1,1) $, since $-(-i)(-i)=1$.

Applying this rule to  electromagnetic tensors, we get
\begin{eqnarray}
&& F^{\tilde{\mu} \tilde{\nu}}= \partial^{\tilde{\mu}} A^{\tilde{\nu}}-\partial^{\tilde{\nu}} A^{\tilde{\mu}}
=\begin{pmatrix}
0 & -i E_1 & -i E_2  & -i E_3  \\
i E_1  & 0 & -B_3 & B_2 \\
i E_2  & B_3 &  0  & -B_1 \\
i E_3  & -B_2 & B_1 & 0  &
\end{pmatrix}
=\begin{pmatrix}
0 & -i \mathbf{E}^t \\
i \mathbf{E} & -\mathbf{B}^\#
\end{pmatrix}
\\
&& G^{\tilde{\mu} \tilde{\nu}}=\begin{pmatrix}
0 & -i \mathbf{B}^t  \\
i \mathbf{B}  & \mathbf{E}^\#
\end{pmatrix}.
\end{eqnarray}

\subsection{The~4 $\times$ 4 Representation of Complex Quaternions}

The~basis elements of quaternion, $1, \mathbf{i}, \mathbf{j}, \mathbf{k}$,  can be represented as $4\times 4$ matrices
\begin{eqnarray}
\begin{pmatrix}
1 & 0 & 0 & 0 \\
0 & 1 & 0 & 0 \\
0 & 0 & 1 & 0 \\
0 & 0 & 0 & 1
\end{pmatrix}, \;
\begin{pmatrix}
0 & -1 & 0 & 0 \\
1 & 0 & 0 & 0 \\
0 & 0 & 0 & -1 \\
0 & 0 & 1 & 0
\end{pmatrix}, \;
\begin{pmatrix}
0 & 0 & -1 & 0 \\
0 & 0 & 0 & 1 \\
1 & 0 & 0 & 0 \\
0 & -1 & 0 &
0
\end{pmatrix},  \;
\begin{pmatrix}
0 & 0 & 0 & -1 \\
0 & 0 & -1 & 0 \\
0 & 1 & 0 & 0 \\
1 & 0 & 0 & 0
\end{pmatrix}.
\end{eqnarray}

A quaternion such as $\d{q}=a+b_1 \mathbf{i}+b_2\mathbf{j}+b_3 \mathbf{k}$ can be represented in the~tensor representation

\begin{eqnarray}
T(\d{q})&&=  a \begin{pmatrix}
1 & 0 & 0 & 0 \\
0 & 1 & 0 & 0 \\
0 & 0 & 1 & 0 \\
0 & 0 & 0 & 1
\end{pmatrix}
+b_1 \begin{pmatrix}
0 & -1 & 0 & 0 \\
1 & 0 & 0 & 0 \\
0 & 0 & 0 & -1 \\
0 & 0 & 1 & 0
\end{pmatrix}
+ b_2 \begin{pmatrix}
0 & 0 & -1 & 0 \\
0 & 0 & 0 & 1 \\
1 & 0 & 0 & 0 \\
0 & -1 & 0 & 0
\end{pmatrix}
+b_3 \begin{pmatrix}
0 & 0 & 0 & -1 \\
0 & 0 & -1 & 0 \\
0 & 1 & 0 & 0 \\
1 & 0 & 0 & 0
\end{pmatrix}
\nonumber \\&&=\begin{pmatrix}
a & -b_1 & -b_2 & -b_3 \\
b_1 & a & -b_3 & b_2 \\
b_2 & b_3 & a & -b_1 \\
b_3 & -b_2 & b_1 & a
\end{pmatrix},
\end{eqnarray} 

where $T$ means the~tensor representation.
When $a=0$,  $T(\d{q})$ has a simple form \vspace{3pt}
$\begin{pmatrix}
0 & -\mathbf{b}^t  \\
\mathbf{b}  & \mathbf{b}^\#
\end{pmatrix}$.

For a quaternion field strength $\d{F}=\vec{E} i-\vec{B} =F_1 \mathbf{i}+F_2 \mathbf{j} + F_3 \mathbf{k}$, the~tensor form of  $\d{F}$ is
\begin{equation}
T(\d{F})= F_1 \begin{pmatrix}
0 & -1 & 0 & 0 \\
1 & 0 & 0 & 0 \\
0 & 0 & 0 & -1 \\
0 & 0 & 1 & 0
\end{pmatrix}
+F_2 \begin{pmatrix}
0 & 0 & -1 & 0 \\
0 & 0 & 0 & 1 \\
1 & 0 & 0 & 0 \\
0 & -1 & 0 & 0
\end{pmatrix}
+F_3 \begin{pmatrix}
0 & 0 & 0 & -1 \\
0 & 0 & -1 & 0 \\
0 & 1 & 0 & 0 \\
1 & 0 & 0 & 0
\end{pmatrix}
=\begin{pmatrix}
0 & -\mathbf{F}^t \\
\mathbf{F} & \mathbf{F}^\#
\end{pmatrix},
\end{equation}
where
$\mathbf{F}\equiv i\mathbf{E}-\mathbf{B}$ which is a vector matrix of the~vector $\vec{F}=i \vec{E} -\vec{B}$.
This is eventually identical to $F^{\tilde{\mu} \tilde{\nu}}+ iG^{\tilde{\mu} \tilde{\nu}}$ \cite{girard2007quaternions,sbitnev2018hydrodynamics,sbitnev2019quaternion}.

\subsection{Complex Electromagnetic Tensor and Electromagnetic Laws}

Let us define $\mathfrak{F}$ and its conjugate $\mathfrak{ F}^*$ as
\begin{eqnarray}
\mathfrak{F}=F^{\tilde{\mu} \tilde{\nu}}+iG^{\tilde{\mu} \tilde{\nu}}, \\
\mathfrak{ F}^*=F^{\tilde{\mu} \tilde{\nu}}-iG^{\tilde{\mu} \tilde{\nu}}.
\end{eqnarray}

A few complex tensors can also be defined as follows,
\begin{eqnarray}
&&\mathcal{D} =\begin{pmatrix}
\frac{\partial}{\partial t}, &  -i\boldsymbol{\nabla}^t
\end{pmatrix}
,  \qquad \qquad \qquad \quad
\mathfrak{J} =\begin{pmatrix}
\rho, & i \mathbf{J}^t
\end{pmatrix}, \nonumber
\\
&&\mathfrak{f} =\begin{pmatrix}
\vec{J}\cdot \vec{E},\;\; & i (\rho \mathbf{E} +\mathbf{J}\times \mathbf{B})^t
\end{pmatrix},
\qquad
\mathcal{T}=\begin{pmatrix}
\mathfrak{u} & i \boldsymbol{\mathfrak{p}}^t  \\
i \boldsymbol{\mathfrak{p}} & \mathbb{T}
\end{pmatrix},
\end{eqnarray}
{where  $\mathbf{J}$ is the~vector matrix of $\vec{J}$,
	$(\rho \mathbf{E} +\mathbf{J}\times \mathbf{B})$ is the~vector matrix of $\rho \vec{E} +\vec{J}\times \vec{B}$,  
	$\boldsymbol{\nabla}$ is the~vector matrix of $\nabla$,  and  $\boldsymbol{\mathfrak{p}}$  is the~vector matrix of $\vec{\mathfrak{p}}$.}

Then, the~following tensor relations hold:
\begin{eqnarray}
&&\mathfrak{F} \mathfrak{F}^*=\mathfrak{F}^* \mathfrak{F}, \\
&&\mathcal{T}=\frac{1}{2}\mathfrak{F} \mathfrak{F}^*,     \label{T} \\
&& \mathcal{D} \mathfrak{F}=\mathfrak{J}  \;\;(=\mathcal{D}\mathfrak{F}^*), \label{50} \\
&& \mathcal{D}\mathcal{T}= - \mathfrak{f} =\frac{1}{2}\mathfrak{J}(\mathfrak{F}+\mathfrak{F}^*),  \label{51} \\
&& \frac{1}{2} \mathfrak{F}\mathfrak{F}=\frac{1}{2} \mathfrak{F}^2 = \mathfrak{L} I, \label{tj}\\
&&\mathcal{T} \bar{\mathcal{T}}= \frac{1}{4} \mathfrak{F}^2 \bar{\mathfrak{F}}^2= \mathfrak{m}^2 I,  \label{tm}\\
&&\rm{Eigenvalues}(\mathcal{T})= \pm \mathfrak{m}, \label{m}
\end{eqnarray}
where $I= (1,1,1,1)$ is unit matrix. $\mathfrak{L}$ and $\mathfrak{m}$ are the~complex Lagrangian and the~electromagnetic mass density (Equation~(\ref{LM})).
All relations can be easily verified by simple calculations.
Actually, the~components of $\mathcal{D}$, $\mathfrak{J}$ and $\mathcal{T}$ are equal to the~components \vspace{2pt} of $\partial_{\tilde{\mu}}$, $J^{\tilde{\mu}}$ and $T^{\tilde{\mu} \tilde{\nu}}_{EM}$, where $\partial_\mu$ is the~four-gradient, $J^\mu$ is the~electric current density and $ T^{\mu \nu}_{EM}$ is the~electromagnetic stress--energy tensor defined as
\begin{equation}
\qquad T^{\mu \nu}_{EM} =
\begin{pmatrix}
\mathfrak{u} & \boldsymbol{\mathfrak{p}}^t  \\
\boldsymbol{\mathfrak{p}} & -\mathbb{T}
\end{pmatrix}.
\end{equation}

Those {listed} relations of complex tensors can be verified by using several known tensor relations in electromagnetism {and tilde-spacetime indices}, instead of the~direct calculation. For example, Equation~(\ref{50}), which represents Maxwell's equations,
can be easily verified from $\partial_\mu F^{\mu \nu}=0$ and $\partial_\mu G^{\mu \nu}= J^{\nu}$.

{The~complex electromagnetic stress--energy tensor $\mathcal{T}$ contains the~information about electromagnetic energy density $\mathfrak{u}$, momentum density $\vec{\mathfrak{p}}$ and stress $\mathbb{T}$, as shown by Equations        (\ref{8}) and (\ref{st}).  It~is interesting that $\mathcal{T}$ is linked to the~electromagnetic mass density, as shown in Equations~(\ref{tm}) and~(\ref{m}). Especially,~Equation~(\ref{m}) cannot be simply derived from known relations of electromagnetism.}

By differentiating both sides of the relation in Equation     (\ref{T}), we get
\begin{eqnarray}
\mathcal{D} \mathcal{T}&&=\mathcal{D}(\frac{1}{2}\mathfrak{F}^* \mathfrak{F}) =\frac{1}{2}((\mathcal{D}\mathfrak{F}^*) \mathfrak{F}+(\mathfrak{F}^{*T} \mathcal{D}^{T})^T\mathfrak{F}))   \nonumber \\
&&=\mathcal{D}(\frac{1}{2}\mathfrak{F} \mathfrak{F}^*) =\frac{1}{2}((\mathcal{D}\mathfrak{F}) \mathfrak{F}^*+(\mathfrak{F}^{T} \mathcal{D}^{T})^T\mathfrak{F}^*)),    \label{64}
\end{eqnarray}
since $\partial_a(A^{a b}B_{b c})=(\partial_a A^{a b})B_{b c}=(\partial_a A^{a b})B_{b c} + A^{a b} (\partial_a B_{b c})$.
Substituting Equation  (\ref{50}) into Equation        (\ref{64}) and comparing it with Equation (\ref{51}), 
we further get the~following relations:
\begin{eqnarray}
(\mathfrak{F}^{T} \mathcal{D}^{T})^T\mathfrak{F}^*
= -(\mathfrak{F} \mathcal{D}^{T})^T\mathfrak{F}^*=\mathfrak{J}\mathfrak{F}\;, \\
(\mathfrak{F}^{*T} \mathcal{D}^{T})^T\mathfrak{F}
=-(\mathfrak{F}^{*} \mathcal{D}^{T})^T\mathfrak{F}=\mathfrak{J}\mathfrak{F}^*.
\end{eqnarray}

\section{Relations between Quaternions and~Two-Spinor Formalism}
\vspace{-6pt}
\subsection{The~Correspondence of~Two-Spinor Representations and Quaternion Representations in Electromagnetism}

Let us start with some basic contents of two-spinor formalism { \cite{penrose1960spinor, bain2000coordinate,penrose1984spinors}}.
Mathematically, any null-like spacetime four-vector $X^\mu$ can be described as a composition of two spinors,
\begin{equation}
X^{\mu} = 1/\sqrt{2} \begin{pmatrix}
\xi  &	\eta   \end{pmatrix}  \boldsymbol{\sigma}^{\mu}  \begin{pmatrix}
\bar{\xi}   \\	\bar{\eta}   \end{pmatrix}   = \frac{1}{\sqrt{2}} \psi^A \sigma^{\mu}_{A A'} \bar{\psi}^{A'},
\end{equation}
where $\boldsymbol{\sigma}^{\mu}$ are sigma matrices \vspace{2pt} $(\sigma^0,\sigma^1,\sigma^2,\sigma^3)$, the~components of $\psi^A$ are $\psi^1 =\xi, \psi^2 =\eta$ for proper complex numbers $\xi$ and $\eta$, and $(\psi^A)^\dagger = \bar{\psi}^{A'}$.
It can be rewritten as
\begin{eqnarray}
\frac{1}{\sqrt{2}} X_\mu \sigma^{\quad \mu}_{BB'} = \psi_B \bar{\psi}_{B'}
\end{eqnarray}
by using the~relation \vspace{2pt} $\bar{\sigma}^{\mu C'C}= \varepsilon^{C'B'}\varepsilon^{CB} \sigma_{BB'}^{\quad \mu}$ and $\sigma_{AA'}^{\quad \mu} \bar{\sigma}^{\; BB'}_\mu =2 \delta_A^B \delta_{A'}^{B'}$, {where $\varepsilon^{AB}$, $\varepsilon^{A'B'}$, $\varepsilon_{AB}$, and $\varepsilon_{A'B'}$ are the~$\varepsilon$-spinors whose components \vspace{2pt} are $\varepsilon^{12}=\varepsilon_{12}=+1,   \varepsilon^{21}=\varepsilon_{21}=-1 $ as follows in \cite{penrose1984spinors}}.

We~now define a spinor $X_{AA'}$ as
\begin{eqnarray}
X_{A A'} \equiv \frac{1}{\sqrt{2}} X_\mu \sigma^{\quad \mu}_{AA'},
\label{new75}
\end{eqnarray}
which is equivalent to $X^\mu$.
The~factor,  which connects a four-vector to a corresponding spinor, is called  ``Infeld--van der Waerden symbol'' {\cite{infeld1933wellengleichung}},
\vspace{2pt} such as $\frac{1}{\sqrt{2}}  \sigma^{\quad a}_{AA'}$ in Equation       (\ref{new75}).  It can be generally written as
$g^{\quad a}_{AA'}$.
We~can extend this notation not only to a null-like four-vector but also to any tensors
by multiplying more than one Infeld--van der Waerden symbols: \vspace{2pt}  any tensor such as $T_{abc..} $ with spacetime indices $a,b,c..$ can be written as a spinor $T_{AA'BB'..}$ with \vspace{2pt} spinor indices $ A, A',B, B'..$,  by multiplying  $T_{abc..}$ with $g^{\quad a}_{AA'}$, $g^{\quad a}_{BB'}$.., such as
$T_{AA'BB'..}=T_{ab..} g^{\quad a}_{AA'} g^{\quad a}_{BB'..}$.
This can be simply written as
\begin{eqnarray}
T_{AA'BB'..}=T_{ab..}.
\end{eqnarray}

Any antisymmetric tensor $H_{ab}=H_{AA'BB'}$ can be divided into two parts
\begin{eqnarray}
H_{AA'BB'}=\phi_{AB} \varepsilon_{A'B'} +\varepsilon_{AB} \psi_{A'B'},
\end{eqnarray}
where $\phi_{AB} =\frac{1}{2} H_{ABC'}^{\qquad C'}$ and $\psi_{A'B'} =\frac{1}{2} H_{C\;\;A'B'}^{\;\; C} $ \vspace{2pt}
(unprimed spinor indices and primed spinor indices can be rearranged back and forth).
If $H_{ab}$ is real, then $\psi_{A'B'}=\bar{\phi}_{A'B'}$ and
\begin{eqnarray}H_{ab}=H_{AA'BB'}=\phi_{AB} \varepsilon_{A'B'} +\varepsilon_{AB} \bar{\phi}_{A'B'}.
\end{eqnarray}

Since an electromagnetic field tensor $F_{ab}$ (Equation \eqref{F}) is antisymmetric, it can be written as
\begin{eqnarray} F_{AA'BB'}=\varphi_{AB} \varepsilon_{A'B'} +\varepsilon_{AB} \bar{\varphi}_{A'B'}
\end{eqnarray}
with an appropriate field $\varphi_{AB}$.
There we find closely related electromagnetic relations \cite{penrose1984spinors}: 
\begin{eqnarray}
\nabla_{AA'} \Phi^{A'B}=\varphi_{A}^{\;\;B},  \label{70}
\\\nabla^{A' B} \varphi_B^{\;\;A}= 2 \pi J^{AA'}, \label{71}
\end{eqnarray}
where $\nabla_{AA'}=\partial_a$ (in Minkowski spacetime) is the~four-gradient, $\Phi_{AA'}=\Phi_{a}$ is the~electromagnetic potential and  $J_{AA'}=J_{a} $ is the~charge-current vector.
The~former is the~relation of electromagnetic potentials and strength fields, and the~latter is equivalent to the~two   Maxwell's equations.

Now, we   prove that
\begin{eqnarray}
\varphi_{A}^{\;\;B} = \frac{1}{2} [(-\vec{E}+i\vec{B})\cdot \vec{\sigma} ]_{A}^{\;\;B} =\frac{1}{2} [(-i\vec{E}-\vec{B})\cdot \vec{\sigma}/i]_{A}^{\;\;B},  \label{78}
\\ \bar{\varphi}^{A'}_{\;\;B'} =\frac{1}{2} [(-\vec{E}-i\vec{B})\cdot \vec{\sigma} ]^{A'}_{\;\;B'} =\frac{1}{2} [(-i\vec{E}+\vec{B})\cdot \vec{\sigma}/i]^{A'}_{\;\;B'}. \label{79}
\end{eqnarray}

Since $\nabla_{AA'}=\frac{1}{\sqrt{2}} \sigma^{\quad a}_{AA'}\partial_a $ \vspace{2pt} corresponds to $\bar{\d{d}}=\frac{\partial}{\partial t}+i\nabla$, $\Phi^{B'B}=\frac{1}{\sqrt{2}} \sigma_b^{\;B'B} \Phi^b$ corresponds to $\d{A}=V+ i \vec{A}$, $\varphi_{A}^{\;\;B}$ \vspace{2pt} corresponds to $\bar{\d{F}}$, and $J^{A'A}=\frac{1}{\sqrt{2}} \sigma_a^{\;A'A} J^a$ corresponds to $\d{J}=\rho+ i \vec{J}$, Equations~(\ref{70}) and (\ref{71}) are 
exactly corresponding to quaternion relations in Equation       (\ref{9}) as follows:
\begin{eqnarray}
\nabla_{AA'} \Phi^{A'B}=\varphi_{A}^{\;\;B} \quad \leftrightarrow \quad \d{d} \bar{\d{A}}=\d{F}, \label{74}\\
\nabla^{A' B} \varphi_B^{\;\;A}= 2 \pi J^{AA'} \quad \leftrightarrow \quad \d{d} \bar{\d{F}}=\d{J}. \label{75}
\end{eqnarray}

Our proof starts from manipulating $F_{AA'BB'}$ as
\begin{eqnarray}
F_{AA'BB'}=\frac{1}{2} F_{\mu\nu} \sigma^{\mu}_{AA'} \sigma^{\nu}_{BB'} =\frac{1}{2} F_{\mu\nu} \sigma^{\mu}_{AA'} \bar{\sigma}^{\nu\; C'C} \varepsilon_{C'B'} \varepsilon_{CB}.
\end{eqnarray}

Then,
\begin{eqnarray}
\varphi_{AB}&&=\frac{1}{2}F_{AA'B}^{\qquad A'}
=\frac{1}{2}F_{AA'BB'}\varepsilon^{A'B'}
\nonumber \\&&=\frac{1}{4}F_{\mu\nu} \sigma^{\mu}_{AA'} \bar{\sigma}^{\nu \; C'C} \varepsilon_{C'B'} \varepsilon_{CB}\varepsilon^{A'B'}
=\frac{1}{4}F_{\mu\nu} \; \sigma^{\mu}_{AA'} \bar{\sigma}^{\nu\; A'C}  \varepsilon_{CB}~,
\\
\bar{\varphi}_{A'B'}&&=\frac{1}{2}F_{AA'\;B'}^{\quad A}
=\frac{1}{2}F_{AA'BB'}\varepsilon^{AB} \nonumber
\\&&=\frac{1}{4}F_{\mu\nu} \bar{\sigma}^{\mu C'C} \sigma^{\nu}_{BB'} \varepsilon_{A'C'} \varepsilon_{AC}\varepsilon^{AB}
=\frac{1}{4}F_{\mu\nu} \;\varepsilon_{A'C'} \bar{\sigma}^{\mu C'B} \sigma^{\nu}_{BB'}~.
\end{eqnarray}

Since
\begin{eqnarray}
\sigma^{\mu}_{AA'} \bar{\sigma}^{\nu \; A'C}=
\begin{pmatrix}
\sigma^{0}\sigma^{0} & -\sigma^{0}\sigma^{1} & -\sigma^{0}\sigma^{2} & -\sigma^{0}\sigma^{3} \\
\sigma^{1}\sigma^{0} & -\sigma^{1}\sigma^{1} & -\sigma^{1}\sigma^{2} & -\sigma^{1}\sigma^{3} \\
\sigma^{2}\sigma^{0} & -\sigma^{2}\sigma^{1} & -\sigma^{2}\sigma^{2} & -\sigma^{2}\sigma^{3} \\
\sigma^{3}\sigma^{0} & -\sigma^{3}\sigma^{1} & -\sigma^{3}\sigma^{2} & -\sigma^{3}\sigma^{3}
\end{pmatrix}_{A  {\raisebox{ 8  pt}{$\scriptstyle C$}}  }=
\begin{pmatrix}
\sigma^{0} & -\sigma^{1} & -\sigma^{2} & -\sigma^{3} \\
\sigma^{1} & -\sigma^{0} & -i\sigma^{3}  & i\sigma^{2}  \\
\sigma^{2} & i\sigma^{3} &- \sigma^{0} & -i\sigma^{1} \\
\sigma^{3} & - i\sigma^{2} & i\sigma^{1} & -\sigma^{0}
\end{pmatrix}_{A  {\raisebox{ 8  pt}{$\scriptstyle C$}}}, \label{80}
\end{eqnarray}
$\varphi_{A}^{\;\;\; D}= \varepsilon^{DB}\varphi_{AB} $ becomes
\begin{eqnarray}
\varphi_{A}^{\;\;\; D}\;&&=\frac{1}{4} F_{\mu\nu} \sigma^{\mu}_{AA'} \bar{\sigma}^{\nu \; A'D} \nonumber \\&&= \frac{1}{4} \; Tr \left [ \begin{pmatrix}
0 & -F_{10} & -F_{20} & -F_{30} \\
F_{10} & 0 & F_{12} & F_{13} \\
F_{20} & -F_{12} & 0 & F_{23} \\
F_{30} & -F_{13} & -F_{23} & 0
\end{pmatrix}
\begin{pmatrix}
\sigma^{0} & -\sigma^{1} & -\sigma^{2} & -\sigma^{3} \\
\sigma^{1} & -\sigma^{0} & -i\sigma^{3}  & i\sigma^{2}  \\
\sigma^{2} & i\sigma^{3} &- \sigma^{0} & -i\sigma^{1} \\
\sigma^{3} & - i\sigma^{2} & i\sigma^{1} &- \sigma^{0}
\end{pmatrix} ^T\; \right ]_{A  {\raisebox{  8 pt}{$\scriptstyle D$}}  }\nonumber \\&&
=\frac{1}{2}(F_{i0}\sigma^{i}-\frac{1}{2}i\,\epsilon_{\;\;k}^{ij}F_{ij}\sigma^k )_A^{\;\;D},
\label{86}
\end{eqnarray}
where $i\;,j\;,k$ are the~three-dimensional vector indices, which have the~value 1, 2 or 3, and $\epsilon_{\;\;k}^{ij}$ is $\epsilon_{pqk} \delta_p^i \delta_q^j$ for the~Levi--Civita symbol $\epsilon_{ijk}$. Einstein summation convention is
understood for three-dimensional vector indices $i,j$ and $k$.
Similar to Equations  (\ref{80}) and (\ref{86}),
\begin{eqnarray}
\bar{\sigma}^{\mu C'B} \sigma^{\nu}_{BB'}=
\begin{pmatrix}
\sigma^{0}\sigma^{0} & \sigma^{0}\sigma^{1} & \sigma^{0}\sigma^{2} & \sigma^{0}\sigma^{3} \\
-\sigma^{1}\sigma^{0} & -\sigma^{1}\sigma^{1} & -\sigma^{1}\sigma^{2} & -\sigma^{1}\sigma^{3} \\
-\sigma^{2}\sigma^{0} & -\sigma^{2}\sigma^{1} & -\sigma^{2}\sigma^{2} & -\sigma^{2}\sigma^{3} \\
-\sigma^{3}\sigma^{0} & -\sigma^{3}\sigma^{1} & -\sigma^{3}\sigma^{2} & -\sigma^{3}\sigma^{3}
\end{pmatrix} _{{\raisebox{  8 pt}{$\scriptstyle C'$}B'}}\!\!=\begin{pmatrix}
\sigma^{0} & \sigma^{1} & \sigma^{2} & \sigma^{3} \\
-\sigma^{1} & -\sigma^{0} & -i\sigma^{3}  & i\sigma^{2}  \\
-\sigma^{2} & i\sigma^{3} & -\sigma^{0} & -i\sigma^{1} \\
-\sigma^{3} & - i\sigma^{2} & i\sigma^{1} & -\sigma^{0}
\end{pmatrix}_{{\raisebox{  8 pt}{$\scriptstyle C'$}B'}} \!\!\!,
\end{eqnarray}
\begin{eqnarray}
\bar{\varphi}^{D'}_{\;\;B'}=\varepsilon^{D'A'}\bar{\varphi}_{A'B'}=-\frac{1}{4}F_{\mu\nu} \;\bar{\sigma}^{\mu D'B} \sigma^{\nu}_{BB'}
= \frac{1}{2}(F_{i0}\sigma^{i}+\frac{1}{2}i\, \epsilon_{\;\;k}^{ij} F_{ij}\sigma^k )^{D'}_{\;\; B'}. \label{88}
\end{eqnarray}

Finally, for an electromagnetic tensor $F_{AA'BB'}$, Equations (\ref{78}) and (\ref{79}) hold. %please revise number of these euqation.
Equations (\ref{86}) and (\ref{88}) also show the~link between Equation~(\ref{T}) and the~spinor form of the~electromagnetic energy--stress tensor $ T_{ab}=\frac{1}{2} \varphi_{AB} \bar{\varphi}_{A'B'}$.

\subsection{General Relations of Quaternion and~Two-Spinor Formalism  and the~Equivalence between Quaternion Basis and Minkowski Tetrads \label{RS}}

Generally speaking, all spinors with spinor indices in two-spinor formalism are directly linked to quaternion. Since $\sigma^{\tilde{\mu}}=(\sigma^0,i\sigma^1,  i\sigma^2,  i\sigma^3)$ \vspace{2pt} is isomorphic to quaternion basis $(1,-\mathbf{i},-\mathbf{j},-\mathbf{k})$, $g^{\quad \tilde{a}}_{AA'}=\frac{1}{\sqrt{2}} \sigma^{\quad \tilde{a}}_{AA'}$ are also isomorphic to $\frac{1}{\sqrt{2}}(1,-\mathbf{i},-\mathbf{j},-\mathbf{k})$.
\vspace{2pt} For any spinors with two spinor indices in the~form $X_{AA'}$, it can be rewritten as $X_{AA'}=X_a g^{\quad a}_{AA'} =X_{\tilde{a}} g^{\quad \tilde{a}}_{AA'}$. \vspace{2pt} It means that we can think of all spinors of the~form $X_{AA'}$ to be obtained by multiplying  the~four-vector with $g^{\quad \tilde{a}}_{AA'}$.

Any spinor $\psi^A$ can be represented with spin basis $o^A, \iota^A$ such as \begin{eqnarray}
\psi^A=a \; o^A +b \; \iota^A
\end{eqnarray}
where $o^A, \iota^A$ is normalized so that $o_A  \iota^A=1 $.
It is well known \vspace{2pt} that Minkowski tetrads $(t^a,x^a,y^a,z^a)$, which is a basis of four-vectors, can be constructed from spin basis $o^A,o^{A'}, \iota^A, \iota^{A'}$ \cite{o2003introduction},
\begin{eqnarray}
g_{0}^{\;\;a}\equiv t^a=\frac{1}{\sqrt{2}}(o^A o^{A'}+\iota^A \iota^{A'}),
\\g_{1}^{\;\;a} \equiv x^a=\frac{1}{\sqrt{2}}(o^A \iota^{A'} +\iota^A o^{A'}),
\\ g_{2}^{\;\;a} \equiv y^a=-\frac{i}{\sqrt{2}}(o^A \iota^{A'} -\iota^A o^{A'}),
\\ g_{3}^{\;\;a}\equiv  z^a=\frac{1}{\sqrt{2}}(o^A o^{A'} -\iota^A \iota^{A'}).
\end{eqnarray}

Therefore,
\begin{eqnarray}
K^a=K^{\mathbf{a}} g_{\mathbf{a}}^{\;\;a}=K^0t^a+K^1 x^a+K^2 y^a+K^3 z^a,
\end{eqnarray}
where a bold index, which represents a ``component'', is distinguished from a normal index.
Any spacetime tensor can be divided into components and basis such as $V^a=V^{\mathbf{a}} \delta_\mathbf{a}^a. $
The~component matrix of Minkowski tetrads with respect to the~spin basis is
\begin{eqnarray}
g_{\mathbf{A} \mathbf{A'}}^{\quad a}=(t^a,x^a,y^a,z^z)=\frac{1}{\sqrt{2}} (\sigma^0,\sigma^1,\sigma^2,\sigma^3)=\frac{1}{\sqrt{2}}\sigma_{\mathbf{A} \mathbf{A'}}^{\quad a}. \label{94}
\end{eqnarray}

We~can replace $g^a$ by the~tilde-tetrads $g^{\tilde{a}}$. The~component matrix of tilde-tetrads $g^{\tilde{a}}$ with respect to the~spin basis is
\begin{eqnarray}
g_{\mathbf{A} \mathbf{A'}}^{\quad \tilde{a}}&&=(t^a,ix^a,iy^a,iz^z) =\frac{1}{\sqrt{2}}(\sigma^0,i\sigma^1,  i\sigma^2,  i\sigma^3 ), \label{94}
\end{eqnarray}
which is isomorphic to $\frac{1}{\sqrt{2}}(\hat{1},-\mathbf{i},-\mathbf{j},-\mathbf{k})$. From this isomorphism, we can set $g_{\mathbf{A} \mathbf{A'}}^{\quad \tilde{a}} (=g_{\mathbf{a}}^{\;\; \tilde{a}})=\frac{1}{\sqrt{2}}(\hat{1},-\mathbf{i},-\mathbf{j},-\mathbf{k})$, which is equivalent to
\begin{eqnarray}
\hat{1} \equiv g_{0}^{\;\; \tilde{a}}=t^a,     \quad  \mathbf{i} \equiv -g_{1}^{\;\;\tilde{a}}=-ix^a,     \quad \mathbf{j} \equiv - g_{2}^{\;\;\tilde{a}}=- iy^a,   \quad \mathbf{k} \equiv- g_{3}^{\;\;\tilde{a}}=-iz^a.  
\end{eqnarray}

Then, any four-vector with tilde-spacetime index can be written as
\begin{eqnarray}
K_a&&=K_{AA'}=K_{\tilde{a}} g^{\quad \tilde{a}}_{AA'}=K_0\hat{1}+(i K_1 ) \mathbf{i}+(i K_2 ) \mathbf{j}+(i K_3 ) \mathbf{k},\\
K^{\tilde{a}}&&=K^{\mathbf{a}} g_{\mathbf{a}}^{\;\;\tilde{a}}=K^0\hat{1}-K^1 \mathbf{i}-K^2 \mathbf{j}-K^3 \mathbf{k},
\end{eqnarray}
where $K^0=K_0,\;K^1=-K_1,\;K^2=-K_2,K^3=-K_3$.

\section{Discussion on Meaning of the~Quaternion and The~Extended Algebra}
\vspace{-6pt}
\subsection{The~Role of Sigma Matrices and Quaternion Basis as Operators}

Let us multiply one of sigma matrices with a tilde-spacetime index by $g^{\tilde{a}}$ as an operator:
Multiplying $(\sigma^{\tilde{1}})^{\mathbf{A'}}_{ \;\;\; \mathbf{B'}}=\begin{pmatrix} 0 &i \\i &0 \end{pmatrix}$ by $g_{\mathbf{A} \mathbf{A'}}^{\quad \tilde{a}}$, we then get
\begin{eqnarray}
g_{\mathbf{A} \mathbf{A'}}^{\quad \tilde{a}} (\sigma^{\tilde{1}})^{\mathbf{A'}}_{ \;\;\; \mathbf{B'}} =\frac{1}{\sqrt{2}}\sigma_{\mathbf{A} \mathbf{A'}}^{\quad \tilde{a}}  (\sigma^{\tilde{1}})^{\mathbf{A'}}_{ \;\;\; \mathbf{B'}}
=\frac{1}{\sqrt{2}}(\hat{1},-\mathbf{i},-\mathbf{j},-\mathbf{k}) (-\mathbf{i}) =\frac{1}{\sqrt{2}}(-\mathbf{i},-1,{-\mathbf{k}},\mathbf{j}). \label{98}
\end{eqnarray}

This is the~operation of changing the~spin basis as
\begin{eqnarray}
o^{A'} \rightarrow i\iota^{A'},\quad
\iota^{A'} \rightarrow  io^{A'}.
\end{eqnarray}

Since $\sigma^{\tilde{\mu}}=(\sigma^0,\sigma^1 i,\sigma^2 i,\sigma^3i)$ \vspace{2pt} is isomorphic to $(1,-\mathbf{i},-\mathbf{j},-\mathbf{k})$,  $\;g_{\mathbf{A} \mathbf{A'}}^{\quad \tilde{a}} (\sigma^{\tilde{1}})^{\mathbf{A'}}_{ \;\;\; \mathbf{B'}}$ can be written as $-g^{\tilde{a}}\mathbf{i}$.
Multiplying $(\sigma^{\tilde{3}})^{\mathbf{A'}}_{ \;\;\; \mathbf{B'}}=\begin{pmatrix} i & 0\\0 &-i \end{pmatrix}$ by $g_{\mathbf{A} \mathbf{A'}}^{\quad \tilde{a}}$, we can see
\begin{eqnarray}
g_{\mathbf{A} \mathbf{A'}}^{\quad \tilde{a}}  (\sigma^{\tilde{3}})^{\mathbf{A'}}_{ \;\;\; \mathbf{B'}} =\frac{1}{\sqrt{2}}\sigma_{\mathbf{A} \mathbf{A'}}^{\quad \tilde{a}} (\sigma^{\tilde{3}})^{\mathbf{A'}}_{ \;\;\; \mathbf{B'}}=\frac{1}{\sqrt{2}}(\hat{1},-\mathbf{i},-\mathbf{j},-\mathbf{k}) (-\mathbf{k}) =\frac{1}{\sqrt{2}}(-\mathbf{k},-\mathbf{j},\mathbf{i},-\hat{1}).
\end{eqnarray}

This corresponds to changing the~spin basis as
\begin{eqnarray}
o^{A'} \rightarrow io^{A'},\quad
\iota^{A'} \rightarrow  -i\iota^{A'},
\end{eqnarray}
and can be written as $-g^{\tilde{a}}\mathbf{k}$.
Multiplying $(\sigma^{\tilde{2}})^{\mathbf{A'}}_{ \;\;\; \mathbf{B'}}=\begin{pmatrix} 0 & 1\\-1 &0 \end{pmatrix}$ by $g_{\mathbf{A} \mathbf{A'}}^{\quad \tilde{a}}$ gives
\begin{eqnarray}
g_{\mathbf{A} \mathbf{A'}}^{\quad \tilde{a}}  (\sigma^{\tilde{2}})^{\mathbf{A'}}_{ \;\;\; \mathbf{B'}} =\frac{1}{\sqrt{2}}\sigma_{\mathbf{A} \mathbf{A'}}^{\quad \tilde{a}} (\sigma^{\tilde{2}})^{\mathbf{A'}}_{ \;\;\; \mathbf{B'}}=\frac{1}{\sqrt{2}}(\hat{1},-\mathbf{i},-\mathbf{j},-\mathbf{k}) (-\mathbf{j}) =\frac{1}{\sqrt{2}}(-\mathbf{j},\mathbf{k},{-\hat{1},-\mathbf{i}}).
\end{eqnarray}

This corresponds to changing the~spin basis as
\begin{eqnarray}
o^{A'} \rightarrow \iota^{A'},\quad
\iota^{A'} \rightarrow  -o^{A'}
\end{eqnarray}
and can be written as $-g^{\tilde{a}}\mathbf{j}$.

Since the~component of $(\sigma^{\tilde{2}})^{\mathbf{A'}}_{ \;\;\; \mathbf{B'}}$ is equal to $\varepsilon^{A'B'}$ and
\begin{eqnarray}
\sigma^{\tilde{\mu} \;\; B'}_{\;A}=\sigma^{\tilde{\mu}}_{AA'}\varepsilon^{A'B}
=(\hat{1},-\mathbf{i},-\mathbf{j},-\mathbf{k})\mathbb{\varepsilon}
=(-\mathbf{j},\mathbf{k},-\hat{1},-\mathbf{i}),
\\
\sigma^{\tilde{\mu} B}_{\quad A'}=\varepsilon^{BA}\sigma^{\tilde{\mu}}_{AA'}
=\mathbb{\varepsilon}(\hat{1},-\mathbf{i},-\mathbf{j},-\mathbf{k})
=(-\mathbf{j},-\mathbf{k},-\hat{1},\mathbf{i}),
\end{eqnarray}
we can interpret that raising or lowering indices means changing spacetime basis.

\begin{figure}[h]
	\centering
	\includegraphics[width=1.00\linewidth]{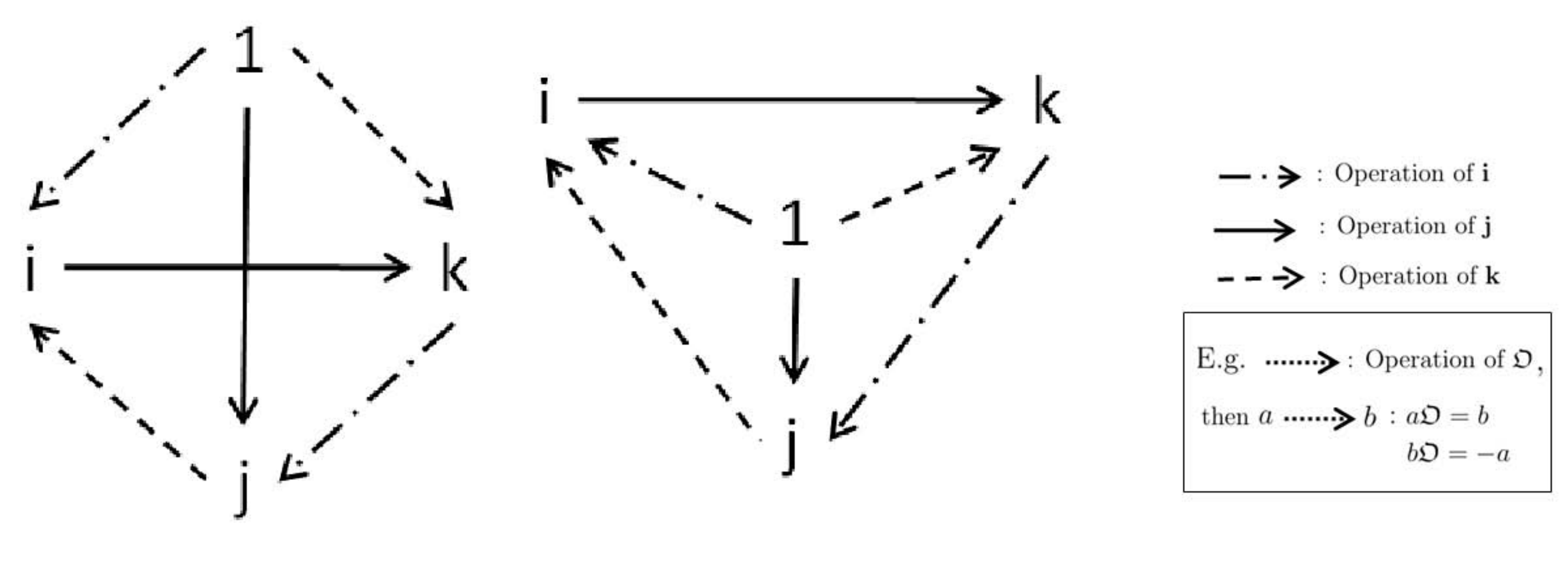}
	\caption{Two figures showing the~result of using the~quaternion basis as an operator set: only the~array of elements is different.}
	\label{fig:production}
\end{figure}

In summary, the~quaternion basis roles as a basis of spacetime itself as well as works as an operator of changing spacetime and spin bases. Similar to the~fact that quantities in classical physics act as operators in quantum mechanics, they allow us to think that spacetime might be formed from fundamental operators.
The~operation on each element of quaternion basis is graphically shown in Figure \ref{fig:production}.
In the~figure, the~three types of arrows indicate the~operations of $\mathbf{i}$, $\mathbf{j}$, and $\mathbf{k}$, respectively. As~shown in the~lower right box, if the~arrow corresponds to the~operation  $\mathfrak{O}$  from $a$ to $b$, then $a \mathfrak{O}=b$ and $b \mathfrak{O}=-a$. As an example, the~solid line indicates the~operation of $\mathbf{j}$, then $\mathbf{i} (\mathbf{j}) = \mathbf{k}$ and $\mathbf{k} (\mathbf{j}) = -\mathbf{i}$.

\subsection{General Discussion on Extended Complex Algebra, and Appropriate Meaning}

Quaternion algebra $\mathbb{H}$ is isomorphic to $\mathbb{C} \times \mathbb{C}$ with non-commutative multiplication rule, and the~elements of $\mathbb{H}$ can be \vspace{2pt} represented with the~secondary complex number $j$ \cite{cowles2017cayley}.  The~set of elements of the~form $q=a+bi+(c+di)j=z_1+z_2j $ where $i^2=j^2=-1, ij=-ji$ is isomorphic to the~set of quaternions $ \d{q}=a+b\mathbf{i}+c\mathbf{j}+d\mathbf{k}$.
In a similar way, \vspace{2pt} we can construct a larger algebraic system of quaternions, which is called ``Octonion'' $\mathbb{O}$ by introducing tertiary complex \vspace{2pt} number $l$, such       as $\mathit{o}=\d{q}_1+\d{q}_2 l $. ``Sedenion'' $\mathbb{S}$, which is an even larger algebraic system than octonion,  can also be derived by performing analogous procedure. This procedure is  called Cayley--Dickson construction.

It is still questionable how octonions and sedenions can be used in physics.
Since octonions have the~similar structure of complex quaternions, they can be used to describe electromagnetism. Furthermore, it is known that a specific octonion is useful to describe SU(3) group, which is the~symmetry group of strong interaction \cite{chanyal2012octonion}.
Sedenion is an algebra which have 16 basis elements. We~suggest that its basis can be written in the~form $q_\mu \otimes q_\nu$, where $q_\mu$ is a quaternion basis $(1,\mathbf{i},\mathbf{j},\mathbf{k})$. We~also speculate that this may be related to SU(4) group, which has 15 generators, or even to the~theory of gravity. Since electromagnetic strength field tensor \vspace{2pt} $F_{ab} = F_{AA'BB'}=\varphi_{AB} \varepsilon_{A'B'} +\varepsilon_{AB} \bar{\varphi}_{A'B'} $  can be expressed in quaternion representation, \vspace{2pt} Weyl tensor  $C_{abcd}=\Psi_{ABCD}\varepsilon_{A'B'}\varepsilon_{C'D'}+\bar{\Psi}_{A'B'C'D'}\varepsilon_{AB}\varepsilon_{CD}$ may be expressed by using sedenion.
The~representation of the~basis and possible uses of each algebraic system are listed in Table~\ref{table1}.

%%%%
\begin{table}[h]

\begin{tabular}{|l|l|l|l|}
			\hline 
			\textbf{Algebraic System} &\textbf{Basis}& \textbf{Products of Basis} &	\textbf{Used}	\\ 
		  \Xhline{1.4pt}
			$\mathbb{C}$		& $1, i$ &       $i^2=-1$
			&   Phase rotation            \\ 
			\hline
			$\mathbb{H} =(\mathbb{C} \times \mathbb{C},* )$& $1, i,j, ij(=\mathbf{k})$ &   \begin{tabular}[c]{@{}l@{}}  $i^2=j^2=-1, ij=-ji$ \\($\mathbf{i}\equiv i,  \mathbf{j}\equiv j, \mathbf{k}\equiv ij $)
			\end{tabular}                        &   \begin{tabular}[c]{@{}l@{}}  Vector rotation and
				\\Lorentz boost, \\electromagnetic laws
			\end{tabular}                                             \\ 
			\hline
			$\mathbb{O} =(\mathbb{C} \times \mathbb{C} \times \mathbb{C},* )$ & \begin{tabular}[c]{@{}l@{}} $1, \mathbf{i}, \mathbf{j}, \mathbf{k}$\\$1l, \mathbf{i}l, \mathbf{j}l, \mathbf{k}l$\end{tabular}
			&\begin{tabular}[c]{@{}l@{}} $l^2=-1$, \\$ e_i= q^\mu l^A =q^\mu_A \quad (l^0=1, l^A=l)$\\
				$(q^\mu\equiv (1,\mathbf{i},\mathbf{j},\mathbf{k}), A=0,1,\mu=0,1,2,3)$\\
				$e_i *e_j(= q^\mu _A*  q^\nu_B)= s_{\mu\nu AB} \;q^\mu q^\nu l^{A+B}$\end{tabular}
			& \begin{tabular}[c]{@{}l@{}}  Rotation of gluon,\\ color charges (SU(3)), \\electromagnetic laws \\with magnetic monopole
			\end{tabular}            \\ 
			\hline
			$\mathbb{S} =(\mathbb{C} \times \mathbb{C} \times \mathbb{C}\times \mathbb{C},* )$
			&\begin{tabular}[c]{@{}l@{}}$ 1, \mathbf{i}, \mathbf{j}, \mathbf{k}$\\$1\mathbf{i}', \mathbf{i}\mathbf{i}',   \mathbf{j}\mathbf{i}', \mathbf{k}\mathbf{i}'$\\ $1\mathbf{j}', \mathbf{i}\mathbf{j}', \mathbf{j}\mathbf{j}', \mathbf{k}\mathbf{j}'$\\$1\mathbf{k}', \mathbf{i}\mathbf{k}', \mathbf{j}\mathbf{k}',\mathbf{k}\mathbf{k}'$
			\end{tabular}                                                   &  \begin{tabular}[c]{@{}l@{}} $e_i=q^{\mu} \otimes q'^{\mu'}=q^{\mu\mu'}$\\
				$(q^\mu=(1,\mathbf{i},\mathbf{j},\mathbf{k}), q'^{\mu'}=(1,\mathbf{i}',\mathbf{j}',\mathbf{k}'))$\\
				$e_i *e_j(=q^{\mu\mu'}*q^{\nu\nu'})= s_{\mu\mu' \nu\nu' } \; q^{\mu\mu'} q^{\nu \nu'}$ \end{tabular}                                                                        & Gravity?, SU(4)?              \\ 
		\hline
	\end{tabular}
		\caption {The~representation of the~basis and possible uses of each algebraic system.
		$s_{\mu\nu AB }, s_{\mu\mu' \nu\nu' }$ are sign operators, which are +1 or $-$1:
		whether $\mu,\nu..$ is 0 or not and whether A and B is 0 or 1 determines the~sign.} 
	\label{table1}
	\centering
\end{table}

We~can think of physical meaning of the~algebras made through Cayley--Dickson construction.
Multiplying complex numbers by a field implies a change in scale and phase of the~field. In this point of view, the~spatial rotation can be interpreted as a kind of two-fold rotation because quaternions can describe three-dimensional spatial rotation and they consist of two independent imaginary units $i$ and $j$. Moreover,    it might be that the~space itself is constructed from a kind of two-fold rotation.
Similarly, since the~basis of octonion can be represented with three complex numbers (one quaternion and one complex number), the~rotation between gluon color charges can be considered as a three-fold rotation. Likewise, if sedenion has useful relation with the~gravity, the~metric of spacetime can be deemed as a four-fold rotation.

\section{Conclusions}

We~have seen that quaternions can describe electromagnetism very concisely and beautifully. They can also represent Lorentz boost and spatial rotation in a simpler way. The~complex conjugation of complex quaternion corresponds to parity inversion of the~physical quantities belonging to the~quaternion. We~can also take a hint from the~$4 \times 4$ matrix representation of quaternion and apply it to define the~complex tensor, which in turn provides a new representation of electromagnetism. We~have verified that the~quaternion representation is directly linked to spinor representation in two-spinor formalism, and then investigated meaning of quaternions; not only as a basis but also as an~operator.

The~use of quaternion could be extended not only for actual calculations, but also to obtain deep insights and new interpretations of physics. Any null-like vectors can be described by two-spinors, and, furthermore, Minkowski tetrads can also be constructed within the two-spinor formalism. This~formalism has the~implication that the~spacetime may come from two-spinor fields. The~beautiful conciseness of quaternion representation of electromagnetism and the~link between quaternion and two-spinor formalism may imply that spinors are the~fundamental ingredients of all fields and the~spacetime  also consists of two-spinor fields. Conversely, if those conjectures are true, then it is natural to explain why the~algebras formed by the~Cayley--Dickson procedure, such       as quaternion, is useful in the~description of~nature.

	\acknowledgments
We would like to thank Gorazd Cvetic for helpful discussions.
This work was supported by the National Research Foundation of Korea
(NRF) grant funded by the Korean government (MSIP) (NRF-2018R1A4A1025334).
\\
\\
%%%%%%%%%%%%%%%%%%%%%%%%%%%%%%%%%%%%%%%%%%
\vspace{6pt} 

%%%%%%%%%%%%%%%%%%%%%%%%%%%%%%%%%%%%%%%%%%
%% optional
%\supplementary{The~following are available online at \linksupplementary{s1}, Figure S1: title, Table S1: title, Video S1: title.}

% Only for the~journal Methods and Protocols:
% \supplementary{The~following are available at \linksupplementary{s1}, Figure S1: title, Table S1: title, Video S1: title. A supporting video article is available at doi: link.}

%%%%%%%%%%%%%%%%%%%%%%%%%%%%%%%%%%%%%%%%%%

%%%%%%%%%%%%%%%%%%%%%%%%%%%%%%%%%%%%%%%%%%
%% optional
%\abbreviations{The~following abbreviations are used in this manuscript:\\
%
%\noindent 
%\begin{tabular}{@{}ll}
%MDPI & Multidisciplinary Digital Publishing Institute\\
%DOAJ & Directory of open access journals\\
%TLA & Three letter acronym\\
%LD & linear dichroism
%\end{tabular}}

%%%%%%%%%%%%%%%%%%%%%%%%%%%%%%%%%%%%%%%%%%
%% optional

\appendix

%\section{}
%\unskip
%\subsection{}

\section{Expansions of a few Quaternion Products in Equation (\ref{9}) \label{appA}}
\vspace{-6pt}
\subsection{Expansions of Products in (3), (4) and (5) of Equation (\ref{9})}
Using the multiplication expression shown in Equation (\ref{2}), the left sides of the relations (3), (4) and (5) in Equation (\ref{9}) are expanded as follows.
\begin{eqnarray}
\d{d} \bar{\d{A}}&&=  (\frac{\partial}{\partial t}-i\nabla)(V-i\vec{A})=(\frac{\partial V}{\partial t} +\nabla \cdot \vec{A})+i\,(-\nabla V-\frac{\partial \vec{A}}{\partial t})-(\nabla \times \vec{A}) \label{dA}\\
\d{d}\bar{\d{F}}&&=(\frac{\partial}{\partial t}-i \nabla)(-i\vec{E}-\vec{B}) \nonumber \\
&&=\nabla \cdot \vec{E} -i\nabla \cdot \vec{B} -(\nabla \times \vec{E}+ \frac{\partial \vec{B}} {\partial t})+i\,(- \frac{\partial \vec{E}}{\partial t} +\nabla \times \vec{B})  \label{dF} \\
\d{d} \bar{\d{J}}&&=  (\frac{\partial}{\partial t}-i\nabla)(\rho-i \vec{J})=(\frac{\partial \rho}{\partial t} +\nabla \cdot \vec{J})+i\,(-\nabla \rho+\frac{\partial \vec{J}}{\partial t})-(\nabla \times \vec{J}) \label{dJ}
\end{eqnarray}

\subsection{The~proof of (7) in Equation (\ref{9}) \label{fd}}
Here we~show that $(\d{F}\d{d})\bar{\d{F}}$ \vspace{2pt} is equal to $\d{F}(\d{d}\bar{\d{F}})$ where $(\d{F}\d{d})$ is the quaternion differential operator.

\begin{eqnarray}
&&(\d{F}\d{d})\bar{\d{F}} \nonumber \\
&&=\left[ (i \vec{E}-\vec{B})(\partial_{t}-i\nabla ) \right] (-\vec{E}i-\vec{B}) \nonumber\\
&&=\left[-\vec{E} \partial_{t} \cdot \vec{E}-(\vec{B} \times \nabla) \cdot \vec{E} -\vec{B} \partial_{t} \cdot \vec{B}+(\vec{E} \times \nabla) \cdot \vec{B} \right] \nonumber\\
\;\;&&+\left[ -i \vec{B} \partial_{t} \cdot E+i (\vec{E} \times \nabla) \cdot \vec{E}+i \vec{E} \partial_{t} \cdot \vec{B}+i (\vec{B} \times \nabla) \cdot \vec{B} \right]\nonumber\\
\;\;&&+\left[\vec{E} \partial_{t} \times \vec{E} -(\vec{B}\cdot \nabla) \vec{E} + (\vec{B} \times \nabla) \times \vec{E} +\vec{B} \partial_{t} \times \vec{B} + (\vec{E}\cdot \nabla) \vec{B}-(\vec{E} \times \nabla) \times \vec{B} \right]\nonumber\\
\;\;&&+\left[i \vec{B} \partial_{t} \times \vec{E}+i (\vec{E}\cdot \nabla) \vec{E}-i (\vec{E} \times \nabla) \times \vec{E}-i \vec{E} \partial_{t} \times \vec{B} + i (\vec{B} \cdot \nabla) \vec{B} -i (\vec{B} \times \nabla) \times \vec{B}\right]\nonumber\\
&&= \vec{J} \cdot \vec{E}+ i \vec{B}\cdot \vec{J} +\rho \vec{B}+ \vec{J} \times \vec{E}+ i (\rho \vec{E} + J \times \vec{B})  \nonumber\\
&&=\d{F}\d{J} ,
\end{eqnarray}

where we have used the following relations:
\begin{eqnarray}
(\vec{B} \times \nabla) \cdot \vec{E} &&= \epsilon_{ijk} B_i \nabla_j \vec{E}_k= \vec{B} \cdot (\nabla \times \vec{E}), \\
\left[ (\vec{B} \times \nabla) \times \vec{E} \right]_i &&= \epsilon_{ipq} (\epsilon_{pjk} B_j \nabla_k) E_q=(\delta_{qj}\delta_{ik}-\delta_{qk}\delta_{ij}) (B_j \nabla_k \vec{E}_q )\nonumber
\\ &&=B_j \nabla_i E_j -B_i(\nabla\cdot \vec{E}),  
\\
\left[\vec{B} \times (\nabla \times \vec{E})\right]_i&&=\epsilon_{iqp} B_q (\epsilon_{pjk} \nabla_j) E_k =  (\delta_{ij}\delta_{qk}-\delta_{ik}\delta_{qj}) B_q \nabla_j E_k \nonumber
\\ &&=  B_j \nabla_i E_j - (\vec{B}\cdot \nabla) E_i,   \label{df}
\\(\vec{B} \times \nabla) \times \vec{E} &&=\vec{B} \times (\nabla \times \vec{E})+(B\cdot \nabla) \vec{E}-\vec{B}(\nabla \cdot \vec{E}).
\end{eqnarray}

\section{The~Proof of Equation        (\ref{26}) \label{ntp}}
\begin{equation}
\nabla \times \mathfrak{p}= \nabla \times (\vec{E}\times \vec{B})= (\vec{E} \cdot \nabla) \vec{B}- (\vec{B} \cdot \nabla) \vec{E} + \vec{E}(\nabla \cdot \vec{B}) -\vec{B}(\nabla \cdot \vec{E}). \label{A9}
\end{equation}
\begin{eqnarray}
\nabla(\vec{E}\cdot \vec{B})&&= \vec{E} \nabla \vec{B}+\vec{B} \nabla \vec{E}  \nonumber\\
&&= \vec{E} \times( \nabla \times \vec{B})+ \vec{B} \times (\nabla \times \vec{E}) + (\vec{E} \cdot \nabla)\vec{B} +(\vec{B} \cdot \nabla)E \;\;\;\text{(from (\ref{df}))}\nonumber\\
&&= \vec{E} \times (\partial_t \vec{E})+\vec{E} \times \vec{J} + \vec{B} \nabla \vec{E} +(\vec{E} \cdot \nabla )\vec{B} \nonumber\\
&&= -\vec{B}\times (\partial_t \vec{B})+ \vec{E} \nabla \vec{B} + (\vec{B} \cdot \nabla) \vec{E}
\end{eqnarray}
where $(\vec{A}\nabla \vec{B})_i\equiv A_j(\nabla_i B_j)$ for vector fields $\vec{A}$ and $\vec{B}$.
Substituting this into Equation        (\ref{A9}), we get Equation        (\ref{26}).

\section{The~Proof of Equation        (\ref{22}) \label{qq}}
As we mentioned on Equation        (\ref{LTQ}), $\d{q}=q_0+i\vec{q}$ is isomorphic to $q_{\mu} \sigma^\mu$ where $q_{\mu}=(q_{0}, \vec{q})=(q_{0}, q_1,q_2,q_3)$, $\sigma^\mu=\{ \sigma^0, \sigma^1, \sigma^2, \sigma^3 \}$ and $g_{\mu \nu}=(1,-1,-1,-1)$.

Let     us introduce some quantities  that are isomorphic to some quaternions
\begin{eqnarray}
&&\partial_\mu \bar{\sigma}^\mu \; \quad \sim \;\;\d{d}=\partial_t-i\nabla \\
&& M_\mu \bar{\sigma}^\mu \quad\sim \;\;\d{F}=i\vec{E}-\vec{B}\\
&& N_\mu\sigma^\mu \quad\sim \;\; \bar{\d{F}}=-i\vec{E}-\vec{B},
\end{eqnarray}
where $\partial_\mu=(\partial_t,\partial_x,\partial_y,\partial_z)$,
$M_{\mu}=(0,-\vec{E} -i \vec{B})$, $N_{\mu}=(0, -\vec{E}+i\vec{B})$.

Then,
\begin{eqnarray}
\d{d}(\bar{\d{F}} \d{F}) &&\sim \partial_\rho \bar{\sigma}^\rho (N_\mu \sigma^\mu M_\nu \bar{\sigma}^\nu)     = (\partial_\rho N_\mu M_\nu) \bar{\sigma}^\rho \sigma^\mu  \bar{\sigma}^\nu
\nonumber \\
&&=(\partial_\rho N_\mu) M_\nu \bar{\sigma}^\rho \sigma^\mu  \bar{\sigma}^\nu+N_\mu (\partial_\rho  M_\nu) \bar{\sigma}^\rho \sigma^\mu  \bar{\sigma}^\nu \nonumber\\
&&=(\partial_\rho N_\mu) M_\nu \bar{\sigma}^\rho \sigma^\mu  \bar{\sigma}^\nu+
N_\mu (\partial_\rho  M_\nu) (2g^{\mu \rho}-\bar{\sigma}^\mu \sigma^\rho)  \bar{\sigma}^\nu \nonumber\\
&&=(\partial_\rho N_\mu) M_\nu \bar{\sigma}^\rho \sigma^\mu  \bar{\sigma}^\nu+
2 (N^\mu \partial_\mu)  M_\nu   \bar{\sigma}^\nu
- N_\mu (\partial_\rho  M_\nu)\bar{\sigma}^\mu \sigma^\rho  \bar{\sigma}^\nu \nonumber\\
&&=(\partial_\rho \bar{\sigma}^\rho N_\mu \sigma^\mu ) M_\nu  \bar{\sigma}^\nu- N_\mu \bar{\sigma}^\mu (\partial_\rho \sigma^\rho  M_\nu \bar{\sigma}^\nu) +
2 (N^\mu \partial_\mu)  M_\nu   \bar{\sigma}^\nu \nonumber\\
&&\sim (\d{d} \bar{\d{F}} ) \d{F}+\bar{\d{F}} (\bar{\d{d}} \d{F}) +2 i (\bar{\d{F}}\cdot \nabla ) \d{F}. 
\end{eqnarray}

We have used the relations $(\bar{\sigma}^\mu \sigma^\nu + \bar{\sigma}^\nu \sigma^\mu)^{A'}_{\;\;B'}=2g^{\mu \nu} \delta^{A'}_{\;\; B'}$ \cite{wess1992supersymmetry} for spinor indices $A,B,A'$ and $B'$.

%\nocite{*}
%\bibliographystyle{ieeetr}
%\bibliography{bibfile}

%%%%%%%%%%%%%%%%%%%%%%%%%%%%%%%%%%%%%%%%%%
% Citations and References in Supplementary files are permitted provided that they also appear in the~reference list here. 

%=====================================
% References, variant A: internal bibliography
%=====================================

%%%%%%%%%%%%%%%%%%%%%%%%%%%%%%%%%%%%%%%%%%
\end{document}